\documentclass[twocolumn]{aastex62}
\usepackage{amsmath}

\graphicspath{{./}{figs/}}

\begin{document}

\title{Temperature, Mass and Turbulence: A Spatially Resolved Multi-Band Non-LTE Analysis of CS in TW~Hya}

\correspondingauthor{Richard Teague}
\email{rteague@umich.edu}

\author[0000-0002-0786-7307]{Richard Teague}
\affil{Department of Astronomy, University of Michigan, 1085 S University Ave., Ann Arbor, MI 48109, USA}
\affil{Max-Planck-Institut f{\"u}r Astronomie, K{\"o}nigstuhl 17, D-69117 Heidelberg, Germany}

\author{Thomas Henning}
\affil{Max-Planck-Institut f{\"u}r Astronomie, K{\"o}nigstuhl 17, D-69117 Heidelberg, Germany}

\author{St\'{e}phane Guilloteau}
\affil{Laboratoire d’astrophysique de Bordeaux, Universit\'{e} de Bordeaux, CNRS, B18N, Alle Geoffroy Saint-Hilaire, 33615 Pessac, France}

\author[0000-0003-4179-6394]{Edwin A. Bergin}
\affil{Department of Astronomy, University of Michigan, 311 West Hall, 1085 S. University Ave, Ann Arbor, MI 48109, USA}

\author[0000-0002-3913-7114]{Dmitry Semenov}
\affil{Max-Planck-Institut f{\"u}r Astronomie, K{\"o}nigstuhl 17, D-69117 Heidelberg, Germany}

\author{Anne Dutrey}
\affil{Laboratoire d’astrophysique de Bordeaux, Universit\'{e} de Bordeaux, CNRS, B18N, Alle Geoffroy Saint-Hilaire, 33615 Pessac, France}

\author{Mario Flock}
\affil{Max-Planck-Institut f{\"u}r Astronomie, K{\"o}nigstuhl 17, D-69117 Heidelberg, Germany}

\author{Uma Gorti}
\affil{SETI Institute/NASA Ames Research Center, Mail Stop 245-3, Moffett Field, CA 94035-1000, USA}

\author[0000-0002-1899-8783]{Tilman Birnstiel}
\affil{University Observatory, Faculty of Physics, Ludwig-Maximilians-Universit\"{a}t M\"{u}nchen, Scheinerstr. 1, D-81679 Munich, Germany}

\begin{abstract}
Observations of multiple rotational transitions from a single molecule allow for unparalleled constraints on the physical conditions of the emitting region. We present an analysis of CS in TW~Hya using the $J=7-6$, $5-4$ and $3-2$ transitions imaged at $\sim 0.5\arcsec$ spatial resolution, resulting in a temperature and column density profile of the CS emission region extending out to 230~au, far beyond previous measurements. In addition, the 15~kHz resolution of the observations and the ability to directly estimate the temperature of the CS emitting gas, allow for one of the most sensitive searches for turbulent broadening in a disk to date. Limits of $v_{\rm turb} \lesssim 0.1 c_s$ can be placed across the entire radius of the disk. We are able to place strict limits of the local H$_2$ density due to the collisional excitations of the observed transitions. From these we find that a minimum disk mass of $3 \times 10^{-4}~M_{\rm sun}$ is required to be consistent with the CS excitation conditions and can uniquely constrain the gas surface density profile in the outer disk.
\end{abstract}

\keywords{astrochemistry -- ISM: molecules -- protoplanetary disks -- techniques: interferometric}

\section{Introduction}

To understand the planet formation process we must first understand the environmental conditions of planetary birthplaces \citep{Mordasini_ea_2012}. Thanks to the unparalleled sensitivity and resolution provided by the Atacama Large (sub-)Millimetre Array (ALMA), we are routinely resolving sub-structures indicative of in-situ planet formation and on-going physical processing \citep{ALMA_ea_2015, Andrews_ea_2016, Perez_ea_2016, Fedele_ea_2017, Dipierro_ea_2018}. Similar features have been observed in high-contrast AO near-infrared imaging tracing the small grain population well coupled to the gas \citep[e.g.][]{vanBoekel_ea_2017, Pohl_ea_2017, Hendler_ea_2018}.

In addition, excess UV emission is interpreted as accretion onto the central star from the disk, with more massive disks accreting at a higher rate \citep{Fang_ea_2004, Manara_ea_2016}. These observations point towards active disks which are able to efficiently redistribute material and angular momentum.

Despite the observational evidence for the redistribution of angular momentum, identifying the physical mechanisms which enables this remains elusive. Two scenarios are likely. Firstly, a turbulent viscosity would be sufficient to transport angular momentum outwards. This is the assumption in the frequently implemented `$\alpha$-visocity' disk model of \citet{Shakura_ea_1973}. Although agnostic about the main driver of the turbulence, this model links the turbulent motions to the viscosity of the disk. Alternatively, angular momentum can be removed through winds \citep{Bai_2017}, evidence of which has been observed in several young sources although is lacking in more evolved counterparts. 

The magneto-rotational instability has been a leading contender as the source for turbulence \citep{Balbus_Hawley_1998, Fromang_Nelson_2006, Simon_ea_2013, Simon_ea_2015, Bai_2015, Flock_ea_2015, Flock_ea_2017}. However estimates of the local ionization rate close to the disk midplane have suggested that there would be insufficient coupling between the rotating gas and the magnetic field \citep{Cleeves_ea_2015a}. Additional instabilities have been shown to generate turbulent without the need for ionization such as the vertical shear instability \citep{Nelson_ea_2013, Lin_Youdin_2015}, gravitational instabilities \citep{Gammie_2001, Forgan_ea_2012} baroclinic instabilities \citep{Klahr_Bodenheimer_2003, Lyra_Klahr_2011} and the zombie vortex instability \citep{Marcus_ea_2015, Lesur_Latter_2016}. Distinguishing between these mechanisms requires a direct comparison of the distributions of non-thermal motions observed in a disk and the predicted distribution from simulations \citep[for example:][]{Forgan_ea_2012, Flock_ea_2015, Simon_ea_2015}.

There have been several attempts to detect non-thermal motions in disks through the additional broadening in line emission in the disks of TW~Hya and HD~163296 \citep{Hughes_ea_2011, Guilloteau_ea_2012, Flaherty_ea_2015, Flaherty_ea_2017, Flaherty_ea_2018, Teague_ea_2016}. Although a promising avenue of exploration, this approach is hugely sensitive to the temperature assumed as the Doppler broadening of the lines does not distinguish between the sources of the motions, either thermal or non-thermal. One must make assumptions about the physical structure of the disk in order to break these degeneracies \citep{Simon_ea_2015, Flaherty_ea_2015, Flaherty_ea_2017}. \citet{Teague_ea_2016} attempted to minimize the assumptions made about the disk structure, inferring physical properties directly from the observed spectra and allowing the temperature and turbulent structure to vary throughout the disk. Without the leverage of an assumed model, the constraints on $v_{\rm turb}$ were larger than other attempts, finding $v_{\rm turb} \lesssim 0.3~c_s$ across the radius of the disk. These constraints were limited by assumptions made about the thermalisation of the energy levels, in particular CN emission was shown to be in non-LTE across the outer disk, while for the single CS transition, as the line was optically thin, the degeneracy between column density and temperature could not be broken without assuming an underlying physical structure. 

However, high-resolution observations of the studied disks show substructures traced in mm-continuum, molecular line emission and scattered light \citep{Andrews_ea_2016, Flaherty_ea_2017, Teague_ea_2017, vanBoekel_ea_2017, Monnier_ea_2017}. Such perturbations from a `smooth' disk model could be sufficient to mask any signal from non-thermal motions which require a precise measure of the temperature to a few Kelvin.

In this paper we present new ALMA observations of CS $J=7-6$ and $J=3-2$ transitions in TW~Hya, the nearest protoplanetary disk at $d = 60.1$~pc \citep{Bailer-Jones_ea_2018} with a near face-on inclination of $i \approx 6\degr$. Combined with the previously published $J = 5-4$ observations \citep{Teague_ea_2016}, we are able to fit for the excitation conditions of the molecule, namely the temperature, density, column density and non-thermal broadening component.

\section{Observations}
\label{sec:observations}

\subsection{Data Reduction}

\begin{deluxetable*}{cccccccccc}
\tablecaption{List of Observations \label{tab:observations}}
\tablehead{\colhead{CS Transition} & \colhead{Frequency} & \colhead{$E_{\rm u}$} & \colhead{$\Delta V_{\rm chan}$} & \colhead{Robust}  & \colhead{UV Taper} & \multicolumn2c{Restoring Beam} & \colhead{Image RMS\tablenotemark{a}} & \colhead{Peak $T_B$} \\ 
\colhead{} & \colhead{\footnotesize(GHz)} & \colhead{\footnotesize(K)} & \colhead{\footnotesize(${\rm m~s}^{-1}$)} & \colhead{} & \colhead{\footnotesize(k$\lambda$)} & \colhead{\footnotesize($\arcsec \times \arcsec$)} & \colhead{\footnotesize($\degr$)} & \colhead{$({\rm K})$} & \colhead{$({\rm K})$}} 
\startdata
$J = 3 - 2$ & 146.96904 & 14.11 & 35 & 2.0 & 320 & $0.57 \times 0.50$ & $91.3$ & 0.56 & 10.7 \\
$J = 5 - 4$ & 244.93556 & 35.27 & 19 & 0.5 & - 	 & $0.59 \times 0.47$ & $91.4$ & 0.46 & 11.6 \\
$J = 7 - 6$ & 342.88285 & 65.83 & 14 & 2.0 & 230 & $0.57 \times 0.51$ & $74.9$ & 0.39 & 10.3 \\
\enddata
\tablenotetext{a}{The resulting noise for the azimuthally averaged spectra is reduced by a factor of $\sqrt{N}$ where $N$ is the number of beams averaged over.}
\end{deluxetable*}

\begin{figure*}
\centering
\includegraphics[width=\textwidth]{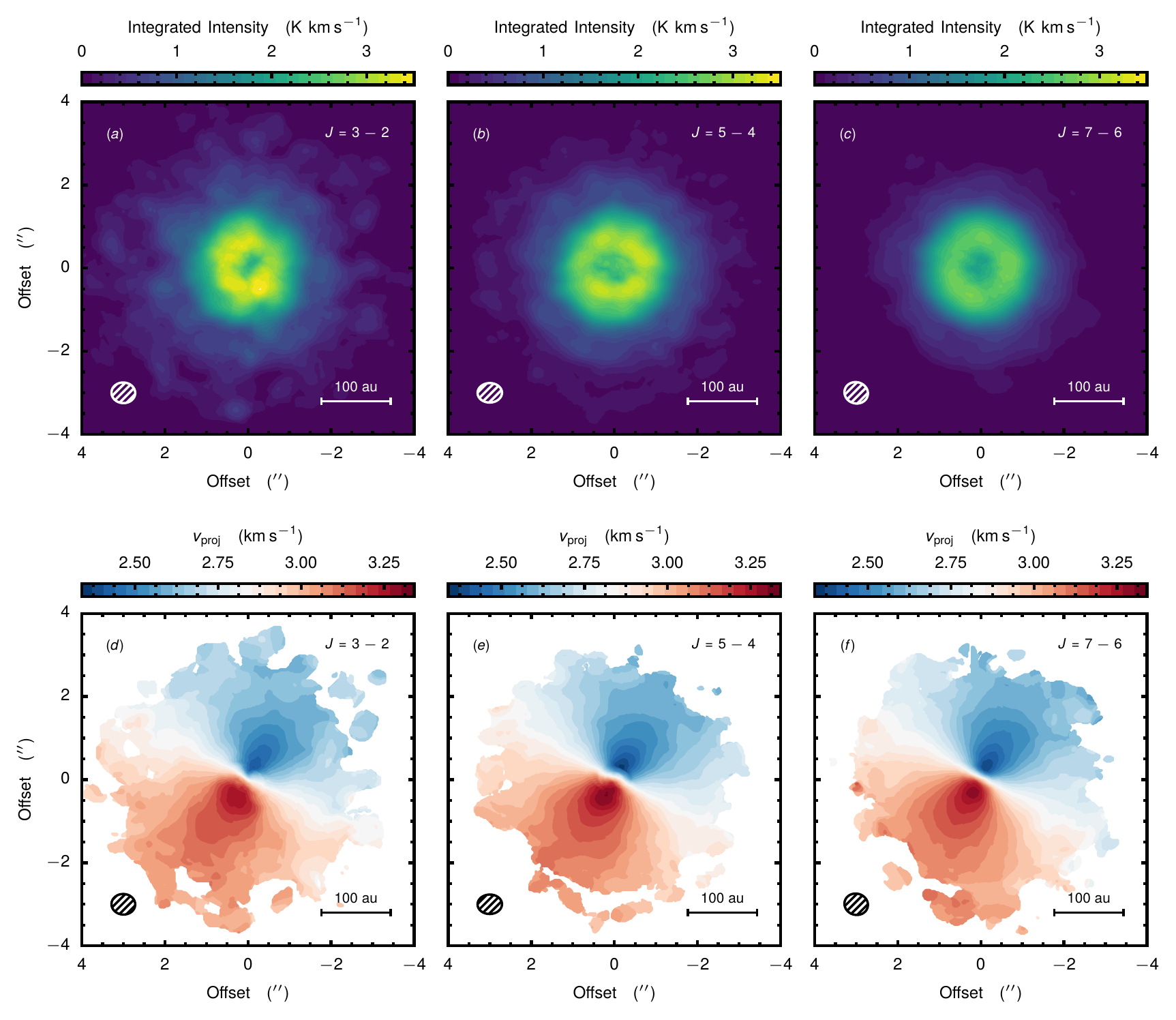}
\caption{Moment maps \added{of the} three CS emission lines: zeroth moment, top, and first moment, bottom. Beam sizes are shown by the hatched ellipse in the bottom left of each panel. \label{fig:moments}}
\end{figure*}

The new observations were part of the ALMA project 2016.1.00440.S, targeting the $(7-6)$ and $(3-2)$ rotational transitions of CS at 342.882850~GHz and 146.969029~GHz in Bands 7 and 4, respectively. Band 6 data comes from project 2013.1.00387.S, originally published in \citet{Teague_ea_2016}.

Band 4 data were taken in October 2016 over the 22$^{\rm nd}$, 25$^{\rm th}$ and 27$^{\rm th}$ with 38, 39 and 40 antennae, respectively with baselines spanning 18.58~m to 1.40~km and a total on-source time of 138.73 minutes. Band 7 data were taken on the 2$^{\rm nd}$ December 2016 utilising 45 antennas with baselines spanning 15.10~m to 0.70~km and a total on-source time of 47.75 minutes. The quasars J1107-449 and J1037-2934 were used as amplitude and phase calibrators for both bands. For each band, the correlator was set up to have a spectral window covering the CS transition with a channel spacing of 15.259~kHz.

The data were initially calibrated with the provided scripts. Phase self-calibration was performed using \texttt{CASA} v4.7. Gain tables were generated on collapsed continuum images then applied to the line spectral windows. Imaging was performed in \texttt{CASA} v5.2. \added{Continuum was subtracted from all line spectral windows using the \texttt{uvcontsub} task. As continuum emission is observed to extend out to $\approx 70$~au \citep{Andrews_ea_2016} this subtraction will not affect the line emission outside this region. Within 70~au, however, regions where the continuum is optically thick may suffer from small absorption effects \citep{Boehler_ea_2017}, however this will be limited to the very inner regions which are not considered in this work.}

All cubes \texttt{CLEAN}ed using a Keplerian mask using parameters consistent with previous observations of TW Hya. To mitigate any differences between observations due to the different beam sizes, we applied a different weighting scheme to each transition. With the least extended baselines, the band 6 data were imaged using a robust weighting using s Briggs parameter of 0.5. To attain a comparable beamsize with the band 4 and 7 data we used natural weighting and included a Gaussian taper to the extended baselines. Each image was centred on the peak of the continuum using the \texttt{fixvis} command, with the centres being consistent within 50~mas in both right ascension and declination. Table~\ref{tab:observations} summarises the observations and resulting images.

The ALMA Technical Handbook quotes a flux calibration uncertainty of 5\% for Band 4 and 10\% for Bands 6 and 7. Comparison of the integrated continuum fluxes for our data with other published values suggest that there are no significant deviations as dicussed in Appendix~\ref{sec:app:fluxcalibration}. In the following we assume a 10\% flux calibration uncertainty for all three bands.

\subsection{Observational Results}

Zeroth and first moment maps were generated using both the Keplerian mask used for cleaning and clipping values below $2\sigma$. These are shown in Figure~\ref{fig:moments}. All three lines display a similar emission pattern with an off-centre peak at $\sim 1\arcsec$ ($60~{\rm au}$) and extending to $\sim 3\arcsec$ ($180~{\rm au}$). There is no clear azimuthal asymmetry within the noise. In addition, all lines exhibit the characteristic pattern of Keplerian rotation.

To derive geometrical properties for the disk we fit a Keplerian rotation pattern, $v_{\rm Kep} = \sqrt{GM_{\star} \, / \, r}$ to the first moment maps. The disk centre, $(x_0,\, y_0)$, inclination $i$, position angle, PA and systemic velocity, $v_{\rm LSR}$, were allowed to vary. Following \citet{Walsh_ea_2017} we convolve the rotation pattern before calculation of the likelihood. The stellar mass was fixed at $0.65~M_{\rm sun}$ \citep{Qi_ea_2004} because the disk inclination is too low to break the degeneracy between $M_{\star}$ and $\sin(i)$ in the velocity pattern.

Results for each line were consistent, yielding an inclination $i = 6.8 \pm 0.1\degr$, ${\rm PA} = 151 \pm 1\degr$ and $v_{\rm LSR} = 2842 \pm 2~{\rm ms^{-1}}$. The uncertainties on these values are dominated by the scatter between the three observations and the statistical uncertainties are an order of magnitude smaller. The derived $M_{\star} \cdot \sin i$ value is consistent with the $0.88~M_{\rm sun}$ and $i = 5\degr$ used to model high resolution CO emission \citep{Huang_ea_2018}. The use of these values would not affect the results in the following analysis.

Following \citet{Teague_ea_2016} \citep[also see][]{Yen_ea_2016, Matra_ea_2017}, we deproject the data spatially and spectrally accounting for the rotation of the disk. Each pixel at deprojected coordinate $(r,\,\theta)$ was shifted by an amount 

\begin{equation}
v_{\rm proj}\,(r,\ \theta) = v_{\rm Kep}(r) \cdot \sin(i) \cdot \cos(\theta),
\end{equation}

\noindent to a common velocity. The data were then binned into annuli with a width of 4~au and then averaged to increase the signal-to-noise ratio. Although this binning is much below the beam size, broader annuli would sample lines arising from significantly different temperatures and thus possessing different line widths. Although in the following text we treat each annulus as independent, in practice they will still be correlated over the size of the beam. 

\begin{figure}
\centering
\includegraphics[width=\columnwidth]{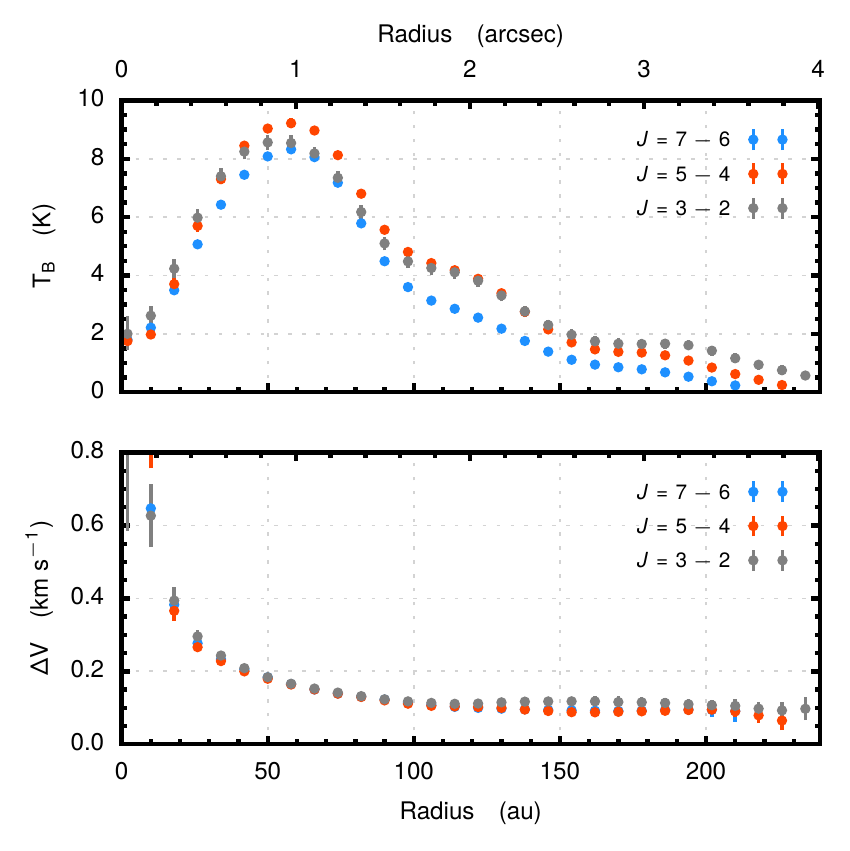}
\caption{Radial profiles of the brightness temperature, $T_B$, and the line width $\Delta V$ derived by fitting a Gaussian profile to the deprojected spectra. All uncertainties are $3\sigma$ statistical uncertainties from the fit. \label{fig:radial_profiles}}
\end{figure}

A Gaussian line profile, characterised by a line width $\Delta V$ and the peak brightness temperature $T_B$, was fit to each azimuthally averaged spectrum \added{(as shown in Appendix~\ref{sec:app:spectra}). A Gaussian profile is a reasonable assumption for such lines which are believed to dominated by thermal broadening. As the CS emission is expected to arise in a narrow layer \citep{Dutrey_ea_2017}, we do not expect significant deviations in the rotation velocity that would lead to non-Gaussian profiles. Furthermore, the face-on orientation of TW~Hya means that the near and far sides of the disk align along the line of sight minimizing the contamination found in more inclined disks \citep[for example HD~163296,][]{Rosenfeld_ea_2013}, suggesting that a Gaussian profile is a reasonable assumption}.

The radial profiles of these parameters are shown in Figure~\ref{fig:radial_profiles}. The broad ring of emission centred at 60~au is clearly seen in the $T_B$ profile, while the knee at 90~au is seen in all three transition. \citet{Teague_ea_2017} argued that this feature was the result of a surface density perturbation. An additional feature in the outer disk, either a drop in $T_{\rm B}$ at $\approx 170$~au or an increase at $\approx 190$~au, is also seen, potentially an outer desorption front from interstellar radiation.

Brightness temperatures peaking at $T_B \approx 9~{\rm K}$ suggests that the lines are optically thin. Assuming that the linewidth in the outer disk is dominated by thermal broadening then we expect temperature of $T_{\rm kin} \gtrsim 25$~K, suggestive of optical depths of $\tau \lesssim 0.4$. This temperature is consistent with the freeze-out temperature of CS, indicating that the CS emission tracing the CS freeze-out surface, comparable to that observed for the optically thin CO isotopologues \citep{Schwarz_ea_2016}.

\section{LTE Analysis}
\label{sec:LTE_Modelling}

\begin{figure*}
\centering
\includegraphics[width=\textwidth]{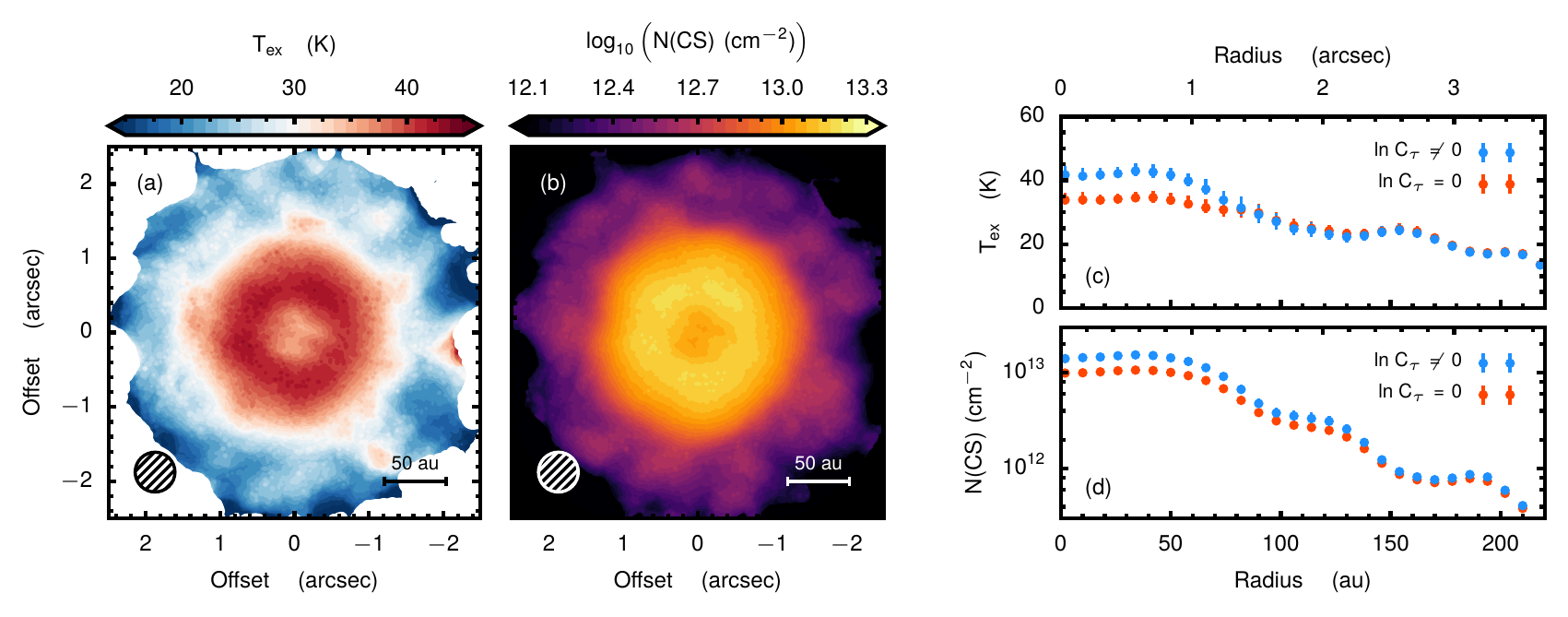}
\caption{LTE analysis for the three CS transitions. Left and centre panels show $T_{\rm ex}$ and $\log_{10} N({\rm CS})$ applied on a pixel-by-pixel basis. The beam in the bottom left shows the average beam of the three transitions. The right two panels show the LTE approach applied to the azimuthally averaged spectra. Blue dots show the results including the optical depth correction, $C_{\tau}$, and red show without. The error bars show the 16th to 84th percentiles of the posterior distributions. \label{fig:LTEfitting}}
\end{figure*}

In this section we assume that the three observed transitions are in Local Thermodynamic Equilibrium (LTE) so that $T_{\rm ex} = T_{\rm kin}$. The excitation temperature will provide a lower limit to gas temperature from the three lines. \citet{Loomis_ea_2018} recently used the same approach to infer the temperature and column density of methyl cyanide, CH$_3$CN, in TW Hya. 

Following \citet{Goldsmith_Langer_1999}, in the optically thin limit the integrated flux can be related to the level population of the upper energy state, 

\begin{equation}
\frac{N_u}{g_u} = \frac{1}{g_u}\frac{8 \pi k \nu^2}{A_{ul} h c^3} \cdot \int T_{\rm B} \, {\rm d}V = \frac{\gamma_u W}{g_u}
\end{equation}

\noindent where $W = \int T_B \, {\rm d} V$ \deleted{and} is the value described by the zeroth moment maps in Figure~\ref{fig:moments}, transformed to units of Kelvin using Planck's law \added{, $N_u$ is the column density of the upper energy level with degeneracy $g_u$, $A_{ul}$ is the Einstein-A coefficient for spontaneous emission and $\nu$ the frequency of the emission}. Under the assumption of LTE we can relate

\begin{equation}
\ln \left( \frac{\gamma_u W}{g_u} \right) = \ln N - \ln Q(T) - \ln C_{\tau} - \frac{E_u}{kT_{\rm ex}},
\end{equation}

\noindent where $N$ is the total column density of the molecule, $Q$ is the partition function, approximated for a linear rotator as,

\begin{equation}
Q(T) = \frac{kT}{hB_0} + \frac{1}{3},
\end{equation}

\noindent and $C_{\tau}$ is the optical depth correction factor, $C_{\tau} = \tau / (1 - \exp(-\tau))$ for a square line profile. In the case of a Gaussian profile this correction is less severe. The optical depth at the line centre is given by

\begin{equation}
\tau = \frac{N_u A_{ul}\, c^3}{8 \pi \Delta V_{\rm FWHM} \nu^3} \left[ \exp\left(\frac{h \nu}{kT_{\rm ex}} \right) - 1 \right]
\end{equation}

\noindent where $\Delta V_{\rm FWHM}$ is the full-width at half-maximum of the line, which can be calculated assuming only thermal broadening, a reasonable assumption given the low levels of turbulence previously reported.

Finally, relating $N_u$ to $N$ through

\begin{equation}
N_u = \frac{g_u N}{Q(T)} \exp \left( \frac{-E_u}{kT_{\rm ex}} \right).
\end{equation}

\noindent allows us fit for $\left\{T_{\rm ex},\ N({\rm CS})\right\}$, self-consistently accounting for possible optical depth effects. As this method considers only the integrated flux, flux calibration uncertainties can be easily considered. We include a systematic flux calibration uncertainty of 10\% in addition to the statistical uncertainty.

As this approach uses only the integrated flux value there are a range of $\left\{T_{\rm ex},\ N({\rm CS})\right\}$ which are consistent with the data but result in highly optically thick lines. We can rule these scenarios out given the $T_B$ profiles shown in Fig~\ref{fig:radial_profiles}: optically thick lines result in $T_B = T_{\rm ex}$, recovering a gas temperature much below the freeze-out temperature of volatile species. To take this into account we include a prior that $\tau < 1$ for all lines.

We use \texttt{emcee} \citep{emcee} to calculate posterior distributions of $T_{\rm ex}$ and $\log_{10}N({\rm CS})$. We assumed uninformative priors,

\begin{equation}
\begin{split}
T_{\rm ex}~({\rm K}) &= \mathcal{U}(10,\ 150) \\
\log_{10}\big( N({\rm CS})~({\rm cm^{-2}}) \big) &= \mathcal{U}(9,\ 14) \\
\tau(T_{\rm ex},\, N({\rm CS})) &= \mathcal{U}(0,\ 1)
\end{split}
\end{equation}

\noindent forcing the models to be optically thin. We used 256 walkers which took 200 steps to burn-in before taking an additional 100 to sample the posterior distribution. The quoted uncertainties are the 16th and 84th percentiles of the posterior distribution, which for a Gaussian distribution represents $1\sigma$ uncertainty.

The results are shown in Figure~\ref{fig:LTEfitting}. The left two panels show the 2D maps of $T_{\rm ex}$ and $\log_{10} N({\rm CS})$ where the fitting was applied on a pixel-by-pixel basis. The two panels on the right show the results when applied to the azimuthally averaged spectra. The blue dots include the correction for optical depth while the gray dots do not. It is only within the inner regions where $\tau$ rises where the correction for the optical depth is appreciable. 

Within the inner 30~au a constant temperature of $T_{\rm ex} \approx 40$~K is found, slowly dropping to 20~K at 200~au. The column density shows two distinct knees at 90~au and 160~au, the former argued for in \citet{Teague_ea_2017} as due to a surface density perturbation. A slight rise in temperature is observed at 150~au. 

The 2D distributions show no significant azimuthal structure in either parameter with deviations being consistent with the uncertainty and show a comparable radial profile to the azimuthally averaged spectra. The 2D maps are only able to achieve a reasonable fit within $\approx 2\arcsec$ because of the noise, demonstrating the strength of the azimuthally averaged approach in tracing the outer disk.

\section{Non-LTE Modelling}
\label{sec:NonLTE_Modelling}

By collapsing our data down to a single integrated flux value as in the previous section, we lose a tremendous amount of information and modelling the entire spectrum allows us to consider a more complex model. In this section we fit the spectra using slab models, allowing us to explore the impact of non-LTE effects and allow us to potentially constrain the local gas density.

We assume that the line profile is well described by an isothermal slab model with

\begin{equation}
T_B(\nu) = \big( J_{\nu}(T_{\rm ex}) - J_{\nu}(T_{\rm bg}) \big) \cdot \left( 1 - e^{-\tau_{v}} \right) + J_{\nu}(T_{\rm bg})
\label{eq:line_profile}
\end{equation}

\noindent where

\begin{equation}
J_{\nu}(T) = \left( \frac{h\nu/k}{\exp(h\nu/kT) - 1} \right),
\end{equation}

\noindent $T_{\rm bg} = 2.73~{\rm K}$ is the background temperature and

\begin{equation}
\tau_{v} = \tau_0 \exp\left(-\frac{(v - v_0)^2}{\Delta V^2}\right),
\label{eq:tau}
\end{equation}

\noindent is the optical depth \citep{Rohlfs_Wilson_1996}. Such a profile accounts for the saturation of the line core at moderate optical depths where the line profile can deviate significantly from a Gaussian profile \citep[see, for example,][]{Teague_ea_2016}. The linewidth is the quadrature sum of thermal broadening and non-thermal broadening components, 

\begin{equation}
\Delta V = \sqrt{\frac{2kT_{\rm kin}}{\mu m_p} + v_{\rm turb}^2}
\end{equation}

\noindent where $\mu = 44$ is the molecular weight of CS and $v_{\rm turb}$ is the line-of-sight velocity dispersion. Non-thermal broadening is parametrised as a fraction of the local sound speed, $\mathcal{M} = v_{\rm turb}\,/\,c_s$ where $c_s$ is the local sound speed. This allows a rough comparison with the frequently used $\alpha$ viscosity parameter via the relation $\alpha \sim \mathcal{M}^2$. However, as this relation is dependent on the form of the viscosity \citep{Cuzzi_ea_2001}, we limit our discussion to only the Mach number.

From a gas kinetic temperature $T_{\rm kin}$, the local gas volume density $n({\rm H}_2)$ and the column density of CS, $N({\rm CS})$, we can calculate the excitation temperatures for each observed level (where $T_{\rm ex} \leq T_{\rm kin}$) and optical depth at the line centre, $\tau_0$. We use the collision rates from \citet{Denis-Alpizar_ea_2018} and assume a thermal H$_2$ ortho-para ratio,

\begin{equation}
\frac{n({\rm ortho-H_2})}{n({\rm para-H_2})} = 9 \times \exp\left(-\frac{170.5~{K}}{T_{\rm kin}}\right)
\end{equation}

\noindent from \citet{Flower_ea_1984a, Flower_ea_1985b}. For a gas of 30~K, this is $\sim 0.03$. These values can then be used to model a full line profile through Equations~\ref{eq:line_profile} and \ref{eq:tau}. 

In practice the excitation calculations are performed with the 0-D code \texttt{RADEX} \citep{vanderTak_ea_2007} assuming a slab geometry, appropriate for the face-on viewing orientation of TW~Hya. To achieve the speed necessary for MCMC sampling, a large grid in $\{T_{\rm kin},\, n({\rm H}_2),\, N({\rm CS}),\, \Delta V\}$ space was run using \texttt{pyradex}\footnote{\url{https://github.com/keflavich/pyradex}}, and linearly interpolated between points. Both $n({\rm H_2})$ and $N({\rm CS})$ were sampled logarithmically while $T_{\rm kin}$ and $\Delta V$ were linearly sampled, with 40 samples along each axis. The resulting grid was checked to confirm that $\tau$ and $T_{\rm ex}$ were smoothly varying over the grid ranges and that linear interpolation was appropriate.

As we are working in the image plane, before comparing the synthetic spectra with data, we must make corrections for the sampling of the ALMA correlator and the imaging process. \citet{Flaherty_ea_2018} discuss the difference in image-plane and $uv$-plane fitting for interferometric data, concluding that for observations with well sampled $uv$-planes, such as the data presented here, there is little difference in the results. Nonetheless, caution must be exercised as spatial correlations could lead to underestimation of the uncertainties.

\begin{figure}
\includegraphics[width=\columnwidth]{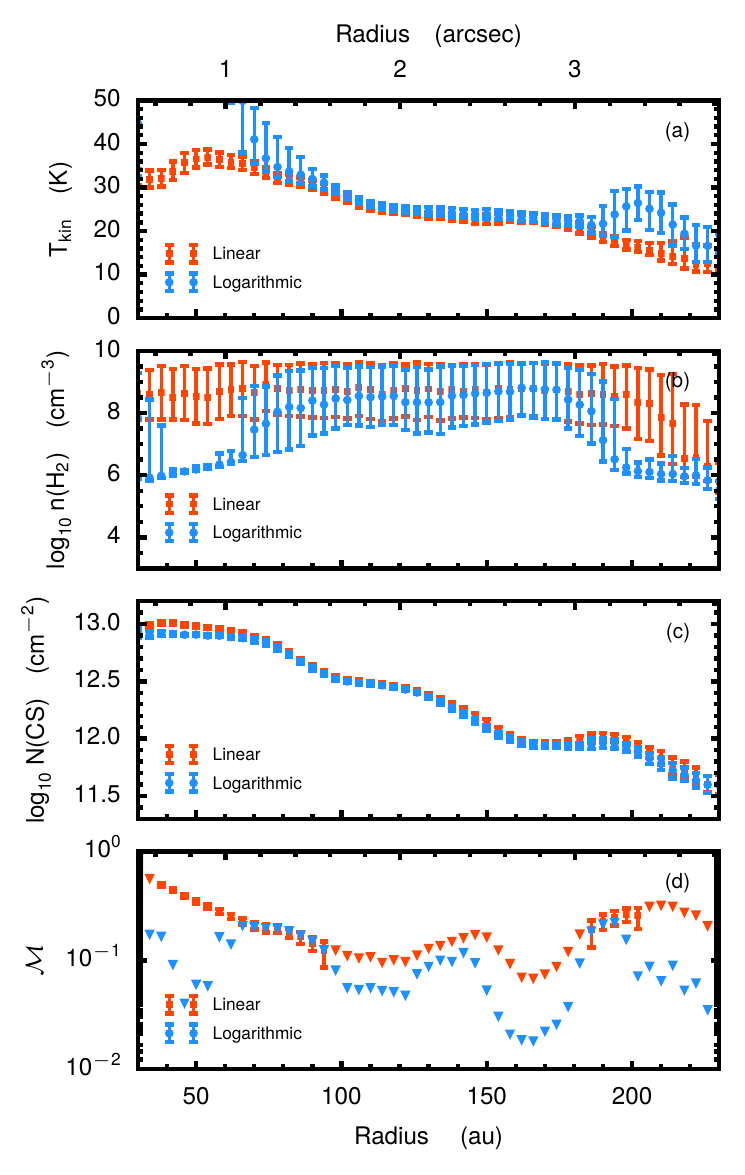}
\caption{Results of the non-LTE fitting. Blue values show the fits to $\log_{10}\mathcal{M}$ while the red points show fits to $\mathcal{M}$. The error bars show the 16th to 84th percentile range of the posterior distribution and the central point is the median. For panel $(d)$, triangles show upper limits for the value, defined as when \replaced{$\mathcal{M}_{16\rm th}$}{the 16th percentile of $\mathcal{M}$} $\leq 10^{-2}$. \label{fig:linex_fit}}
\end{figure}

Spectra are generated at a sampling rate of 150~Hz, a sampling rate of 100 times the observations, before being sampled down to 15~kHz and Hanning smoothed by a kernel with a width 15~kHz. This step is essential as with such narrow line widths this can result in underestimating the intrinsic peak brightness temperature by $\sim 20\%$ and overestimating the width by $\sim 10\%$. Figure~13 from \citet{Rosenfeld_ea_2013} demonstrates how this process affects the emission morphology.

In addition to modelling the emission, we attempt to account for possible spectral correlations in the noise by modelling the noise component using Gaussian Processes, implemented with the Python package \texttt{celerite} \citep{celerite}. \added{A Gaussian Process is a probabilistic non-parametric approach to modelling smoothly varying functions and readily allows for the inclusion of covariances between data points.} The noise was modelled with an approximate Matern-3/2 \replaced{term}{kernel}, \added{an approximation of a Gaussian to mimic the Hanning smoothing applied to the data in the correlator, which is} specified by the amplitude of the noise and the correlation length in velocity, $\sigma_{\rm rms}$ and $\ell$, respectively. \added{The kernel describes how points are correlated over a given dimension, in this case the spectra axis. Figure 1 of \citet{Czekala_ea_2017} provides and example of such kernels.} A simple harmonic oscillator kernel was also tested however no significant difference was found between kernels. These parameters were ultimately considered nuisance parameters, \added{used only to consider how different noise models affected the results}, and marginalised over in the calculation of the excitation condition posterior distributions.

Due to the finite resolution of the data, the synthesized beam will smear out the emission spatially. As discussed in \citet{Teague_ea_2016}, this will lead to a broadening of the lines in the inner region of the disk, while at larger radii, these effects are negligible. We do not attempt to correct for such beam smear effects. The effects are two-dimensional in nature and attempting to model them in one dimension introduces significant uncertainty to the modelling procedure.

In total, for each radial position the three lines can be fully specified by 13 free parameters: the local physical conditions, $\{T_{\rm kin},\, n({\rm H}_2),\, N({\rm CS}),\, \mathcal{M}\}$; the center of each line, $\{v_0\}_i$; and the noise model for each observation $\{\sigma,\,\ell^2\}_i$. All parameters were given uninformative (uniform) priors, ranging across values expected in a protoplanetary disk,

\begin{equation}
\begin{split}
T_{\rm kin}~({\rm K}) &= \mathcal{U}(10,\ 150) \\
\log_{10}\big( n({\rm H_2})~({\rm cm^{-3}}) \big) &= \mathcal{U}(4,\ 10) \\
\log_{10}\big( N({\rm CS})~({\rm cm^{-2}}) \big) &= \mathcal{U}(9,\ 14) \\
\log_{10} \mathcal{M} &= \mathcal{U}(-5,\ 0) \\
\end{split}
\end{equation}

\noindent We additionally run a set of models where we fit for $\mathcal{M}$ rather than $\log_{10}\mathcal{M}$ for which we impose a prior of $\mathcal{M} = \mathcal{U}(0,\ 1)$. Each fit consisted of 256~walkers, each taking 1500~steps for a burn-in period, then an additional 250 to sample the posterior distribution.

The results are shown in Figure~\ref{fig:linex_fit}, with blue points showing the fit to $\log_{10}\mathcal{M}$ and gray points to the linear fit. \added{Example of the covariances between parameters and their posterior distributions can be found in Appendix~\ref{sec:app:nonLTE_fitting}.} The error bars denote the 16th to 84th percentile of the posterior distributions around the median value. Both approaches yield comparable values, consistent with the LTE approach described in Section~\ref{sec:LTE_Modelling}. This is not surprising as high H$_2$ densities are found resulting in thermalisation of the transitions out to $\approx 190$~au.

We see a significant deviation between the models, one where $\mathcal{M}$ is varied linearly and the other where it is varied logarithmically, in the inner 90~au which is due to the artificial broadening of the line from the beam smear. In brief, the broadened lines require either a larger temperature or a large non-thermal broadening component. When fitting for $\log_{10}\mathcal{M}$, changes in the temperature are preferred over changes in the non-thermal broadening due to the temperature only being considered linearly. Conversely, the fit for $\mathcal{M}$ prefers solutions with larger non-thermal broadening without requiring an increase in the temperature. At these higher temperatures assumed in the logarithmic fit, the $J=3-2$ line is considerably less emissive than the $J=7-6$ and thus to maintain a higher $J=3-2$ line flux, the volume density must be artificially decreased to reduce the importance of collisional excitation.

The noise models were able to converge, resulting in Gaussian distributions of their parameters which were not correlated with any other parameter. We have marginalized over these distributions when analysing the distributions of the parameters describing the disk physical conditions. See Appendix~\ref{sec:app:nonLTE_fitting} for an example of the posterior distributions.

\section{Discussion}
\label{sec:discussion}

\subsection{Disk Physical Structure}

\begin{figure}
\includegraphics[width=\columnwidth]{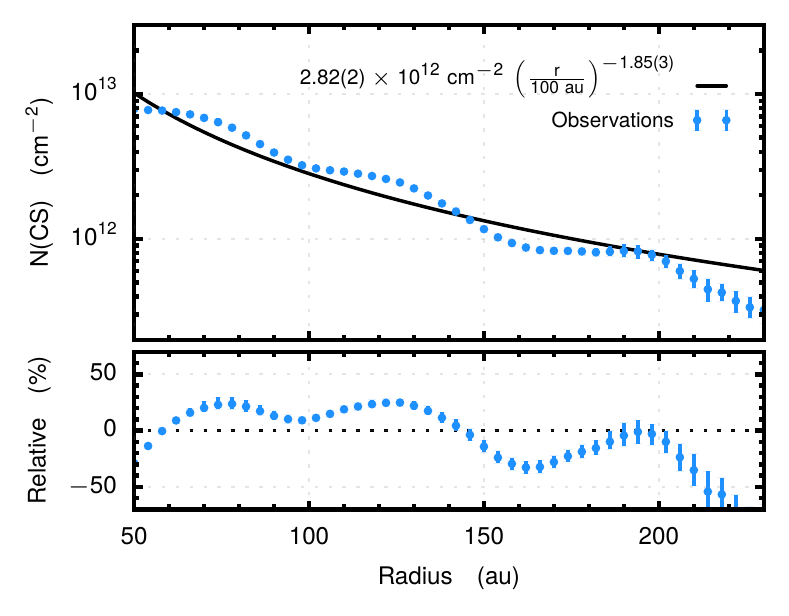}
\caption{Comparing a power-law profile fit to the CS column density in gray to the derived values in blue. The bottom panel shows the relative residual between the two. \label{fig:column_density_comparison}}
\end{figure}

Both LTE and non-LTE approaches paint the same picture: CS is present across the entire extent of the disk in a region which slowly cools from $\approx 40$~K in the inner disk to $\lesssim 20$~K at 200~au.

One interpretation for this is that the CS layer is bounded by the CS `snow surface' \citep{Schwarz_ea_2016, Loomis_ea_2018}. The binding energy of CS is 1900~K resulting in a desorption temperature of $T_{\rm desorb} \approx 31$~K, although dependent on local gas pressures which are unknown, comparable to the temperatures observed for CS. However, a clear boundary between gaseous and ice forms of CS may not be present due to the chemical reprocessing expected on the grain surfaces. Observations of high inclination disks would be able to identify whether the CS emission has a sharp lower boundary.

The column density profiles show two distinct knee features at 90~au and 160~au, both seen in the $T_B$ profiles in Fig.~\ref{fig:radial_profiles}. Figure~\ref{fig:column_density_comparison} compares the column density derived in Section~\ref{sec:NonLTE_Modelling} with a power-law profile fit (shown by the black line). The residuals, shown in the lower panel, shows deviations of up to 20\% in $N({\rm CS})$ at 120 and 160~au. Despite these deviations, models assuming a simple power-law column density, such as those in \citet{Teague_ea_2016}, would be able to adequately model the true column density profile.

The dip at 90~au was previously argued by \citet{Teague_ea_2017} to be due to a significant perturbation in the gas surface density needed to account for a gap traced the scattered light \citep{vanBoekel_ea_2017}. While these results confirm that the emission feature is due to a change in column density rather than temperature, they are unable to distinguish between a local change in CS abundance and a total depletion of gas. Similar features have been observed in high resolution $^{12}$CO observations \citep{Huang_ea_2018}.

At the outer edge of the disk the density drops to a sufficiently low value that, unlike inwards of $\approx 190$~au, the volume density of H$_2$ can be constrained. The apparent constraints inwards of 90~au are, as discussed before, an artefact of the limited angular resolution. Observations of higher frequency lines \added{with higher critical densities} will allow for this method to be sensitive to \added{the} higher densities \added{of the inner disk and allow us to extend these surface density constraints} \deleted{, extending the region where the density is constrained} to smaller radii.

\subsection{Turbulence}

\begin{figure}
\includegraphics[width=\columnwidth]{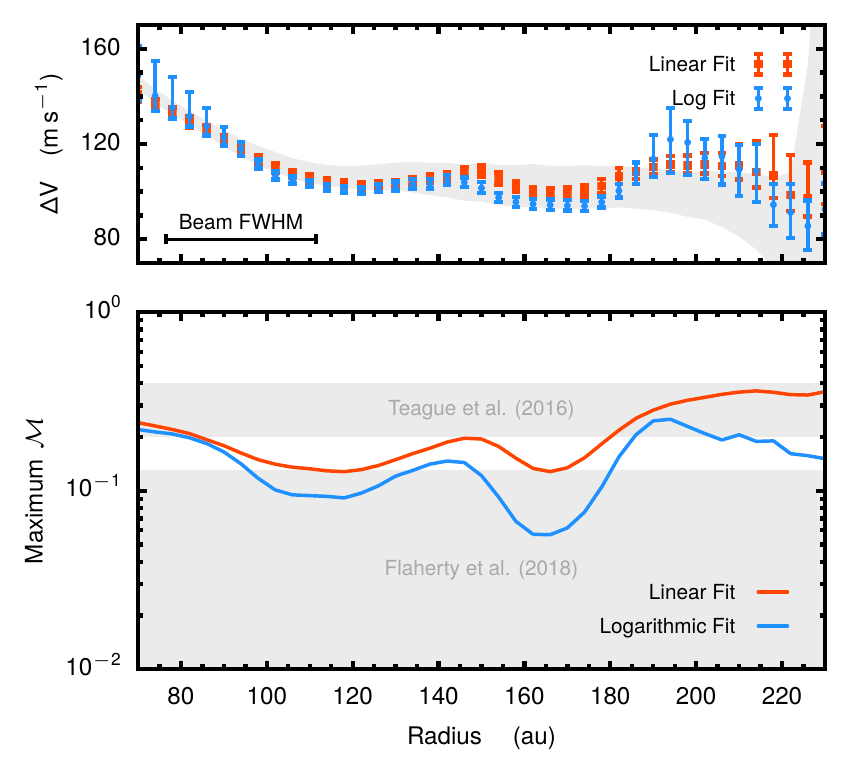}
\caption{Top: Upper limits on $\Delta V$ compared to the observed values which span the gray shaded region. Bottom: $2\sigma$ upper limits on the non-thermal broadening. The shaded regions show the limits from \citet{Teague_ea_2016} and \citet{Flaherty_ea_2018}. \label{fig:turbulence}}
\end{figure}

Determination of the non-thermal broadening requires an accurate measure of the local temperature in order to account for the thermal contribution. By constraining the temperature through multiple transitions minimizes assumptions about the thermal structure and provides \deleted{provides} the most accurate measure of the gas temperature to date. \added{As the CS lines are optically thin, this temperature will be the contribution function-weighted average of the emitting column.} With our derived temperature profile we are therefore able to derive spatially resolved limits on the required non-thermal broadening to be consistent with the data.

For TW~Hya, multiple studies have been already been undertaken \citep{Hughes_ea_2011, Teague_ea_2016, Flaherty_ea_2018}, finding a range of non-thermal broadening values and upper limits, $\mathcal{M} \lesssim 0.4$. As discussed in \citet{Flaherty_ea_2018}, differences in these limits are primarily driven by the different assumptions about the underlying thermal structure and how this couples to the density structure. \citet{Teague_ea_2016} caution, however, that constraints of $\mathcal{M} \lesssim 0.03$ requires constraining the thermal structure to near Kelvin-precision, a limit which is achieved with the data presented in this manuscript.

Under the assumption that the presented three CS lines arise from the same vertical layer in the disk, an assumption which requires observations of edge-on disks to test, we are able to remain agnostic about the thermal and physical structure of the disk. As shown in the bottom panel of Fig.~\ref{fig:turbulence}, we are able to place a $2\sigma$ upper limit out to 230~au. Both approaches (linear and logarithmic fits of $\mathcal{M}$ in red and blue, respectively) yield comparable results. The rise in limits inwards of 100~au is due to the broadening arising due to the beam smearing, while outside 180~au the limits increase due to the lower SNR of the data. The fits yields limits consistent with the values found by \citet{Flaherty_ea_2018}, $\mathcal{M} \leq 0.13$. These are a factor of a few lower than \citet{Teague_ea_2016} due to the warmer temperature derived ($T_{\rm kin} = 28$~K at 100~au compared to 12~K as in \citet{Teague_ea_2016}) as only a single CS transition was available.

Two dips are observed in the profiles at $110$~au and 165~au, consistent with the dips in the column density. The resulting linewidths, plotted in the top panel of Fig.~\ref{fig:turbulence} show that they yield comparable widths to the observations, however over-produce the linewidth outside 200~au, likely due to the lower SNR of the data. We leave interpretation of these features for future work.

Deriving limits for $\mathcal{M}$ using a parametric modelling approach, where physical properties and chemical abundances are described as analytical functions, as in \citet{Guilloteau_ea_2012}, \citet{Flaherty_ea_2015, Flaherty_ea_2017, Flaherty_ea_2018} and \citet{Teague_ea_2016} requires a well constrained molecular distribution which for CS is not known. Furthermore, there is mounting evidence that spatially varying abundances of C and O within the disk can radially alter the local chemistry \citep{Bergin_ea_2016}, further limiting the accuracy of a simple analytical prescription.

\subsection{Minimum Disk Mass}

\begin{figure}
\includegraphics[width=\columnwidth]{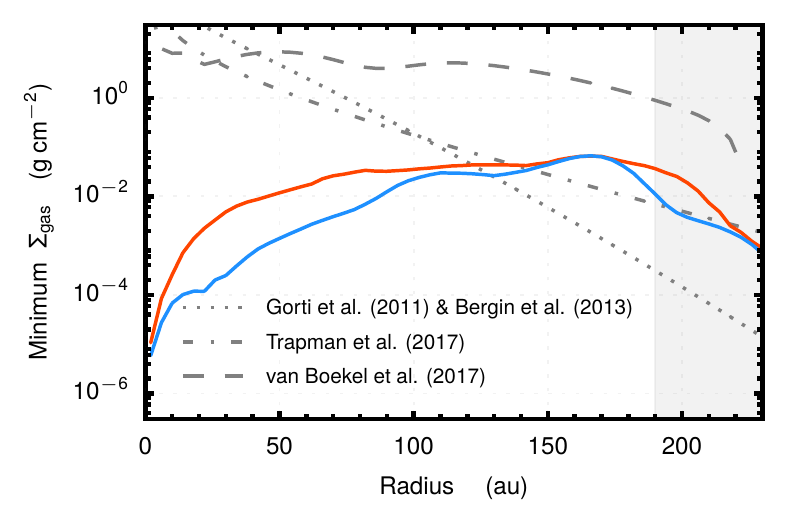}
\caption{Inferred minimum gas surface density and resulting minimum disk masses using the derived $n({\rm H_2})$ values. The blue line shows the results when fitting for $\log_{10}\mathcal{M}$ and the red when fitting for $\mathcal{M}$. The gray shaded region shows where $n({\rm H_2})$ has been constrained. The dotted and dot-dash lines show the surface density profiles used to model the HD emission in \citet{Bergin_ea_2013} \citep[using the surface density from][]{Gorti_ea_2011} and \citet{Trapman_ea_2017}. The dashed line shows the surface density used by \citet{vanBoekel_ea_2017} to model the scattered light emission profile. \label{fig:minimum_disk_mass}}
\end{figure}

\added{Many studies have attempted to measure the mass of the TW~Hya disk through observations of both the mm~continuum and gas emission lines. With differing assumptions these have resulted in a range of masses spanning $5 \times 10^{-4}~M_{\rm sun}$ to $6\times10^{-2}~M_{\rm sun}$ \citep{Weintraub_ea_1989, Calvet_ea_2002, Thi_ea_2010, Gorti_ea_2011, Favre_ea_2013}. Arguably the most accurate approach is to use hydrogen deuteride, HD, as this molecule should be tracing the H$_2$ gas most closely. Modelling of the HD $J=1-0$ transition, \citet{Bergin_ea_2013} concluded that the mass of the disk must be $M_{\rm disk} > 0.05~M_{\rm sun}$. More recently, \citet{Trapman_ea_2017} used additional observations of the $J = 2-1$ transition to find a mass of between $6\times10^{-3}$ and $9\times10^{-3}~M_{\rm sun}$.} This large range is due primarily to the sensitivity of the HD emission to the assumed thermal structure and differences in assumed the cosmic D/H ratio. Models of HD emission show that it is almost entirely confined to the warm inner disk, $r < 100$~au, where the gas is warm enough to sufficient excite the fundamental transition. Although this region accounts for almost all the disk gas mass, the HD flux is insensitive to the cold gas reservoir at smaller radii and is thus a minimum disk mass.

As we have limits on the requied H$_2$ density as a function of radius in Section~\ref{sec:NonLTE_Modelling}, we are able to place a limit on the minimum gas surface density and thus disk mass needed to recover the inferred excitation conditions. It is important to note that this technique does not require the assumption of a molecular abundance, such as those using HD, but rather constrains the H$_2$ gas directly through collisional excitation. It therefore provides an excellent comparison for techniques which aim to reproduce emission profiles and provides a unique constraint for surface densities at large radii in the disk.

To scale a midplane density to a column density we assume a Gaussian vertical density structure, 

\begin{equation}
\rho_{\rm gas} = \frac{\Sigma_{\rm gas}}{\sqrt{2 \pi} H_{\rm mid}} \times \exp \left( - \frac{z^2}{H_{\rm mid}^2} \right)
\end{equation}

\noindent where we take the pressure scale height, 

\begin{equation}
H_{\rm mid} = \sqrt{\frac{k T_{\rm mid} r^3}{\mu m_H G M_{\star}}},
\end{equation}

\noindent which is dependent on the assumed midplane temperature, $T_{\rm mid}$. Observations of the edge-on Flying Saucer have shown CS emission to arise from $\lesssim 1 H_{\rm mid}$ \citep{Dutrey_ea_2017}. The angular resolution of these observations do not allow for distinction between the case of two elevated, thin molecular layers at $\pm 1 H_{\rm mid}$ or a continuous distribution below $H_{\rm mid}$. Measurements of the midplane temperature estimate this to be 5 -- 7~K for the mm dust \citep{Guilloteau_ea_2016} and $\approx 12$~K for the gas \citep{Dutrey_ea_2017}. As we find $T_{\rm kin} = 20$ -- 35~K, this suggests that CS is not tracing the midplane, but rather a slightly elevated region, so that we would overestimate the pressure scale height and underestimate the midplane density. In combination, these uncertainties should mitigate one another allowing for a first-order estimation of the minimum surface density.

For this estimate we use both fits from Section~\ref{sec:NonLTE_Modelling} for the minimum $n({\rm H_2})$ to infer a minimum $\Sigma_{\rm gas}$, shown in Figure~\ref{fig:minimum_disk_mass}. Integrating these minimum surface densities we find an average minimum disk mass of $3\times10^{-4}~M_{\rm sun}$, fully consistent with the estimates from HD emission. Observations of transitions which thermalise at higher densities would extend the sensitivity of this approach such that it can distinguish between models predicting different disk masses.

The shaded region at $r > 190$~au highlights the region where $n({\rm H_2})$ was measured rather than a lower limit constrained. In this region we expect the plotted minimum surface densities to be close to the $\Sigma_{\rm gas}$ value rather than just a minimum value, however the accuracy will be limited by the assumptions made above the vertical structure of the disk. Nonetheless, these profiles provide unique constraints on the gas surface density in the outer regions and are highly complimentary to studies using optically thin CO isotopologues which trace $\Sigma_{\rm gas}$ within the CO snowline \citep{Schwarz_ea_2016, Zhang_ea_2017}.

In Figure~\ref{fig:minimum_disk_mass} we additionally plot surface density profiles from \citet{Gorti_ea_2011}, used in \citet{Bergin_ea_2013} to model the HD flux, the best-fit profile from \citet{Trapman_ea_2017}, also used to model the HD flux and finally the profile from \citet{vanBoekel_ea_2017} used to model scattered light emission. From our lower limit we are able to rule out the model from \citet{Gorti_ea_2011} which contains insufficient material in the outer disk to recover the excitation conditions required by the CS transitions. The profile from \citet{Trapman_ea_2017} is broadly consistent with the lower limits, however would likely not suffice if the H$_2$ densities inwards of 190~au were constrained and would likely become inconsistent when the height of the CS emission surface was taken into account. Therefore the model from \citet{vanBoekel_ea_2017} provides the most consistent profile in the outer disk. This is not surprising as this profile was found by fitting the radial profile of scattered light out to $\sim 200$~au, while the previous two surface density profiles were inferred from integrated flux values.

Although uncertainties in $M_{\star}$, $H_{\rm mid}$, $H_2$ ortho to para ratios and the location of the emission will propagate into the minimum mass, these are negligible compared to the lack in sensitivity of this approach to $n_{\rm H_2} \gtrsim 10^7~{\rm cm^{-3}}$ due to the thermalisation of the $J=7-6$ transition. For commonly assumed power-law surface density profiles, 95\% of the disk mass is within 80~au for TW~Hya, far interior to where this technique is sensitive. Observations of higher density tracers, such as the $J = 9-7$ transition of CS would enable tighter constraints of $\Sigma_{\rm gas}$ at a larger range of radii.

\section{Summary}

We have used spatially and spectrally resolved observations of the $J=7-6$, $5-4$ and $3-2$ transitions of CS in TW Hya to constrain the physical conditions from where CS arises both in a spatially resolved manner and as a radial profile. Accounting for the rotation of the gas and azimuthally averaging the spectra allows us to apply the latter method out to 230~au.

Through both a LTE and a non-LTE approach to fitting the line ratios we find an azimuthally symmetric physical structure. This approach demonstrated that the transitions were thermalized and thus provided a lower limit to the local H$_2$ density. The column density of CS was found to have two significant knees at 90 and 160~au, however distinguishing between an abundance depletion of a true depletion in the total gas surface density cannot be made. We are able to place an upper limit on the non-thermal broadening component for all three lines \added{in regions traced by the CS emission, $z \lesssim H$}, finding $\mathcal{M} \lesssim 0.1$ across the disk, consistent with previous determinations for this source \citep{Hughes_ea_2011, Flaherty_ea_2018}.

Extrapolating the H$_2$ volume density limits to a minimum gas surface density profile allowed us to place a limit on the minimum mass of the disk of $3 \times 10^{-4}~M_{\rm sun}$, in line with constraints from \added{molecular line emission including} HD emission \citep{Bergin_ea_2013, Trapman_ea_2017}, in addition to constraining the gas surface density profile in the outer disk. Observations of higher $J$ transitions will extend the sensitivity of this method to larger densities and thus allow for tighter constraints on the disk mass.

This paper serves as a template for future multi-band observations and demonstrates the power of line excitation analyses in determining spatially resolved physical structures.

\acknowledgments
We thank the anonymous referee for comments which have improved the clarity of this paper. This paper makes use of the following ALMA data: \\ADS/JAO.ALMA\#2016.1.00440.S and \\ADS/JAO.ALMA\#2013.1.00387.S. ALMA is a partnership of ESO (representing its member states), NSF (USA) and NINS (Japan), together with NRC (Canada), NSC and ASIAA (Taiwan), and KASI (Republic of Korea), in cooperation with the Republic of Chile. The Joint ALMA Observatory is operated by ESO, AUI/NRAO and NAOJ. This work was supported by funding from NSF grants AST- 1514670 and NASA NNX16AB48G. RT would like to thank Ryan Loomis with helpful discussion and Adam Ginsburg for his extensive help with \texttt{pyradex}.

\appendix

\section{Flux Calibration}
\label{sec:app:fluxcalibration}

\begin{figure}
\centering
\includegraphics[]{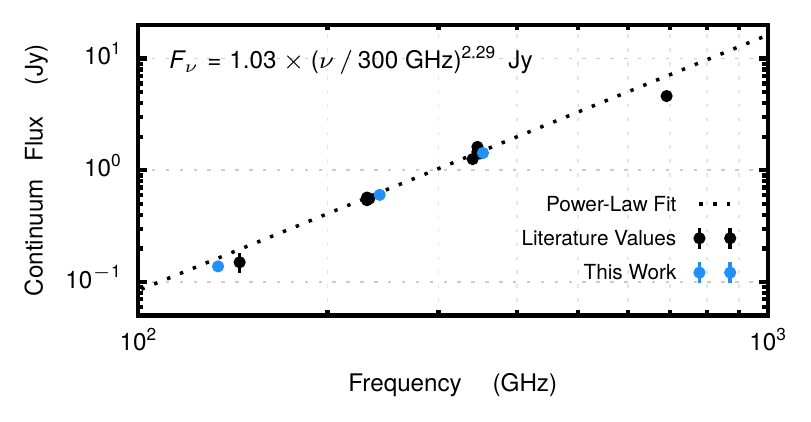}
\caption{Continuum flux measurements for TW~Hya at multiple wavelengths. The blue points show measurements from the data present in this paper while black are literature values. \label{fig:app:fluxcalibration}}
\end{figure}

To check for any significant differences which may arise in the absolute flux scaling of the data, we compare the continuum flux values measured in each of our bands, 0.138, 0.603 and 1.430~Jy at 134, 242 and 353~GHZ, respectively, with previously published data. We plot these in Figure~\ref{fig:app:fluxcalibration} using flux measurements from \citet{Qi_ea_2004, Qi_ea_2006, Hughes_ea_2009, Tsukagoshi_ea_2016} and \citet{Zhang_ea_2016}.

No significant deviation is found relative to the literature values suggesting the flux calibration was performed adequately. 

\section{Azimuthally Averaged Spectra}
\label{sec:app:spectra}

\begin{figure}
\centering
\includegraphics[]{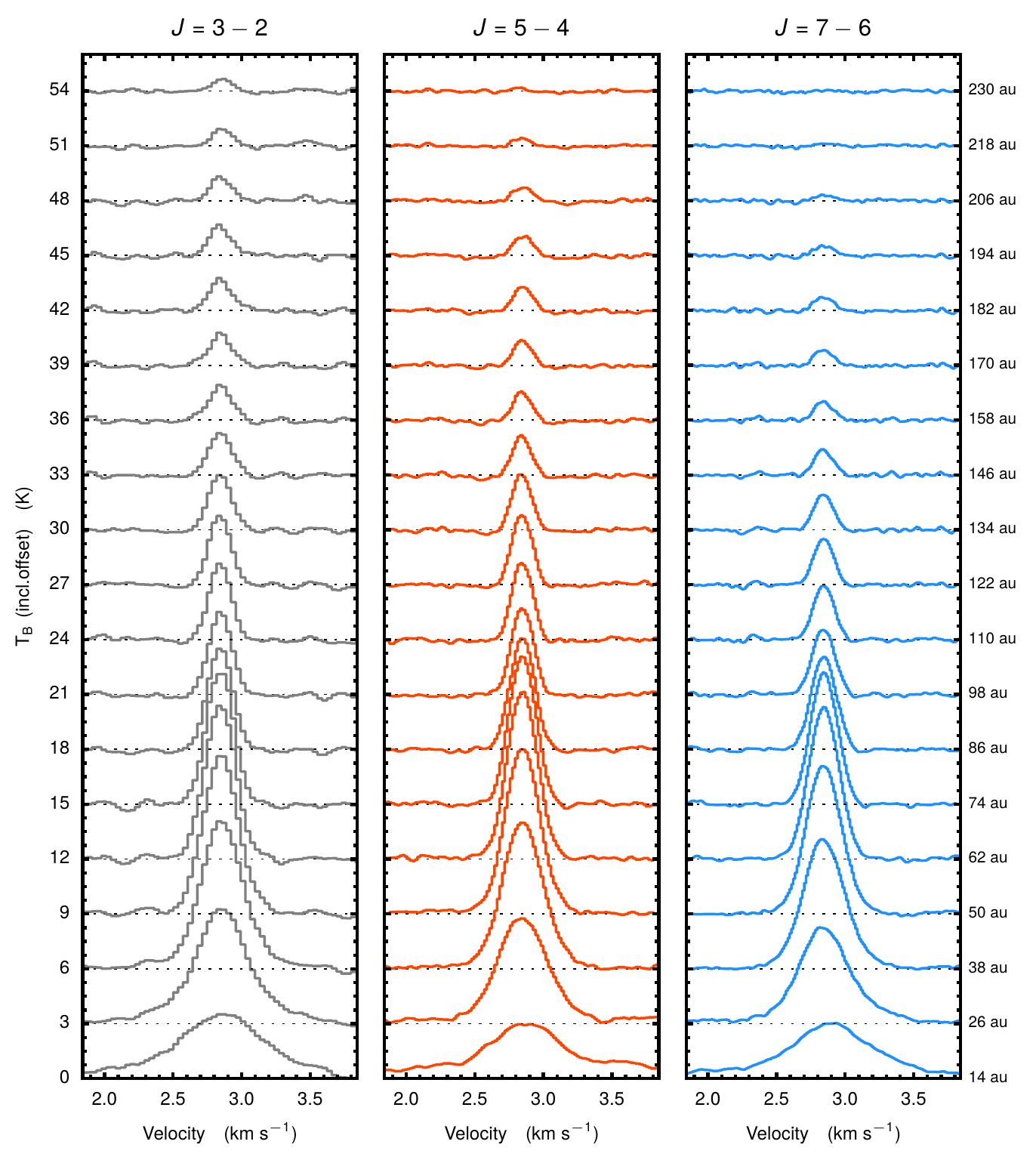}
\caption{Azimuthally averaged spectra centred on $V_{\rm LSR} = 2.842~{\rm km\,s^{-1}}$. The label on the right shows the radius. There is a 3~K offset between each radius. \label{fig:app:spectra}}
\end{figure}

Figure~\ref{fig:app:spectra} shows the azimuthally averaged spectra for the three transitions. Within the inner $\approx 25$~au the beam smear results in highly non-Gaussian line profiles, however, outside this radius a Gaussian profile is an apt description. We detect $J = 3 - 2$ emission out to 230~au, $J = 5 - 4$ out to 220~au and $J = 7 - 6$ out to 210~au. These radii are consistent with the outer edge of $^{12}$CO at 215~au \citep{Huang_ea_2018}.

\begin{figure}
\centering
\includegraphics[]{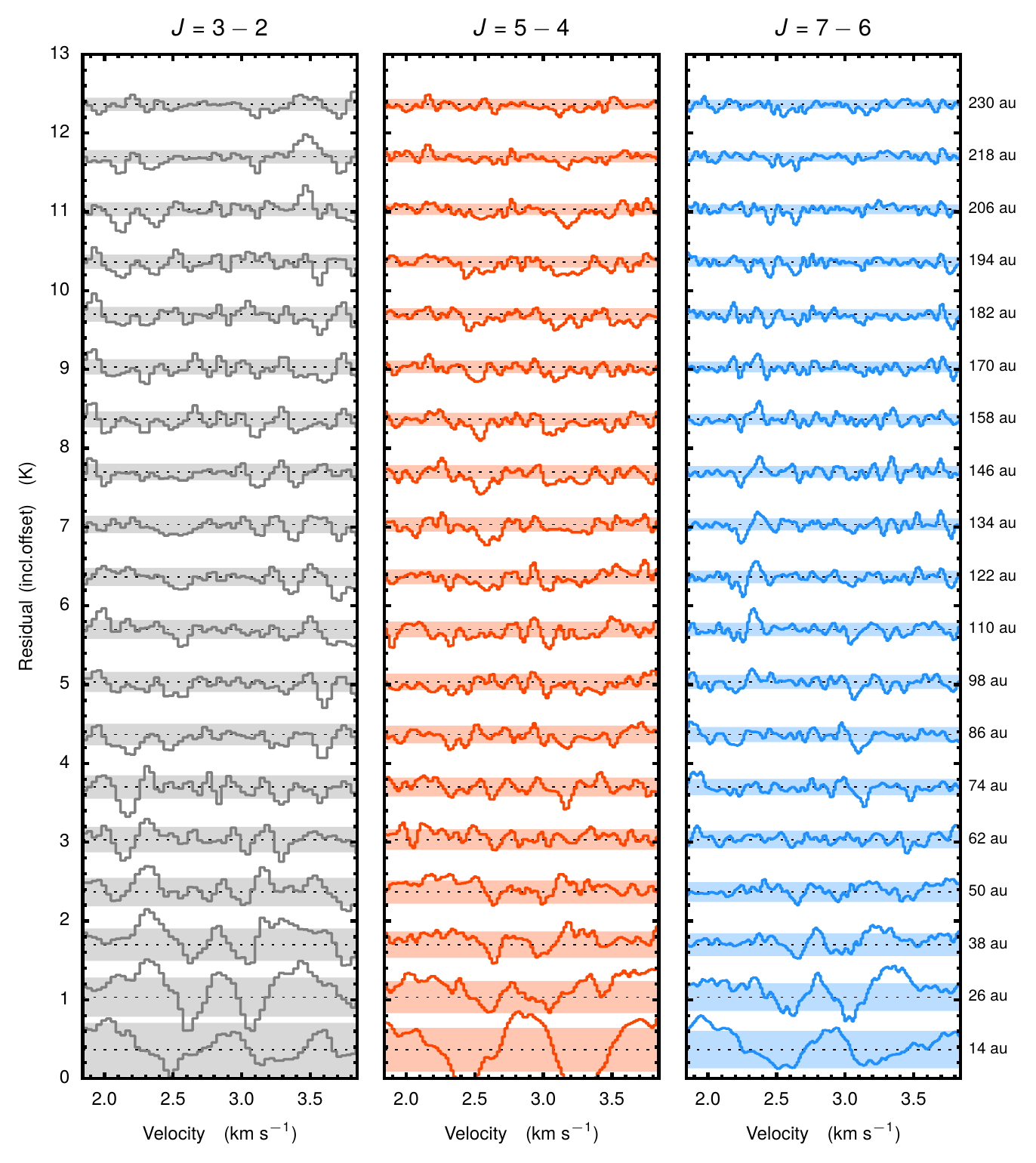}
\caption{Residuals from the azimuthally averaged spectra shown in Fig.~\ref{fig:app:spectra} and a Gaussian line profile. The shading behind each residual shows the noise for that radius taking into account the stacking of independent spectra. \label{fig:app:spectra_residuals}}
\end{figure}

\added{In Fig.~\ref{fig:app:spectra_residuals} we show the residuals from a Gaussian fit to the data. We calculate the noise at each radius taking into account the number of beams averaged over to produce the spectrum. While at smaller radii there appear to be large systematic deviations from a Gaussian profile due to the beam smearing, they are of comparable magnitude to the noise and should thus not significantly bias the results.}

\section{Posterior Distributions for Non-LTE Modelling}
\label{sec:app:nonLTE_fitting}

\begin{figure}
\centering
\includegraphics[]{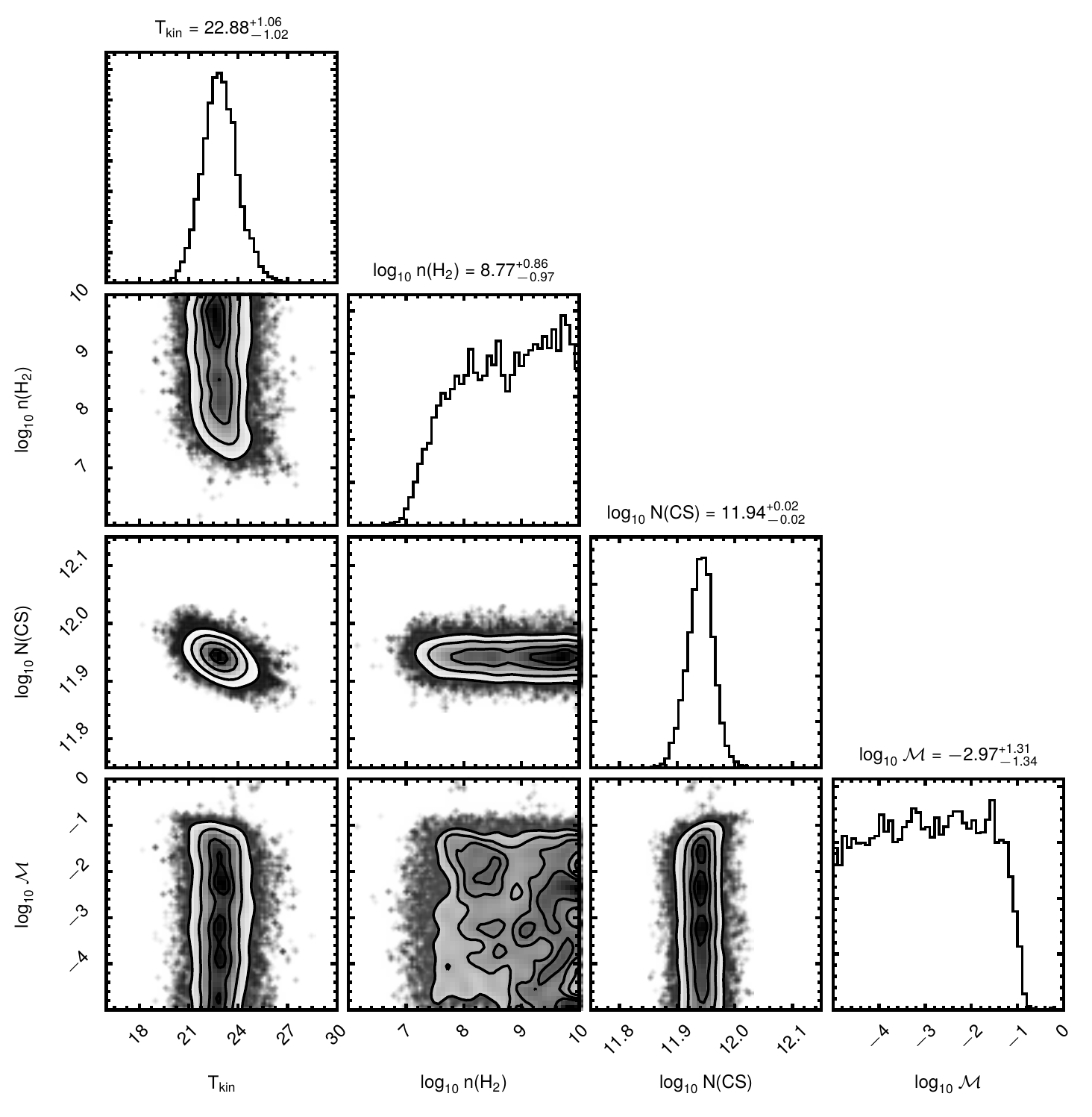}
\caption{Posterior distributions of the excitation conditions from a representative fit at 170~au when fitting with $\log_{10} \mathcal{M}$. \label{fig:app:corner_physical_log}}
\end{figure}

In this Appendix we discuss the MCMC sampling of the posterior distributions  used in the non-LTE fitting described in Section~\ref{sec:NonLTE_Modelling}. We take a representative selection at 170~au. Figures~\ref{fig:app:corner_physical_log} and \ref{fig:app:corner_noise} show the posterior distributions of the excitation conditions and noise models, respectively. At the top of each column of panels is the median value of the distribution with the uncertainties corresponding to the 16th and 84th percentile of the distribution.

Figure~\ref{fig:app:corner_physical_log} shows that both $T_{\rm kin}$ and $N({\rm CS})$ are well constrained and only slightly correlated, while only limits can be placed on $n({\rm H_2})$ and $\log_{10}\mathcal{M}$. The steep fall-off of the $\log_{10}\mathcal{M}$ posterior distribution demonstrates that a tight upper limit has been found.

For the noise properties, no correlation between the parameters are observed. $\sigma$, with units of Kelvin, and demonstrate the significant increase in SNR achieved through the azimuthal averaging compared to the channel noise reported in Table~\ref{tab:observations}. The correlation length, $\ell$, are in units of ${\rm km\,s^{-1}}$, yielding $\ell \, / \, \Delta V$ values of between 3 and 5. This is consistent with the noise seen in Figure~\ref{fig:app:spectra} where noise features appear correlated over 3 to 5 channels.

\begin{figure}
\centering
\includegraphics[]{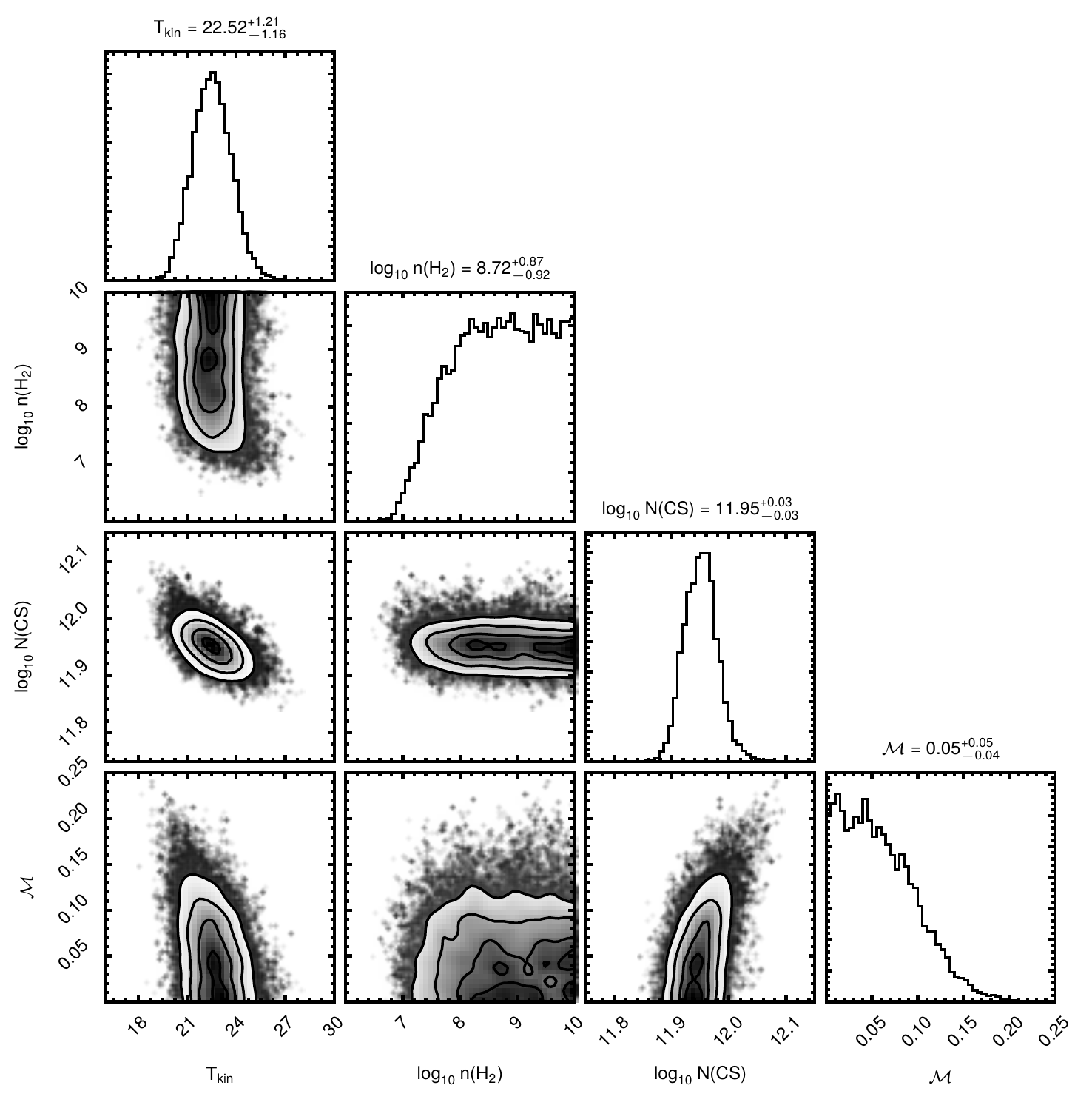}
\caption{Posterior distributions of the excitation conditions from a representative fit at 170~au when fitting with a linear $\mathcal{M}$. \label{fig:app:corner_physical_lin}}
\end{figure}

\begin{figure}
\centering
\includegraphics[]{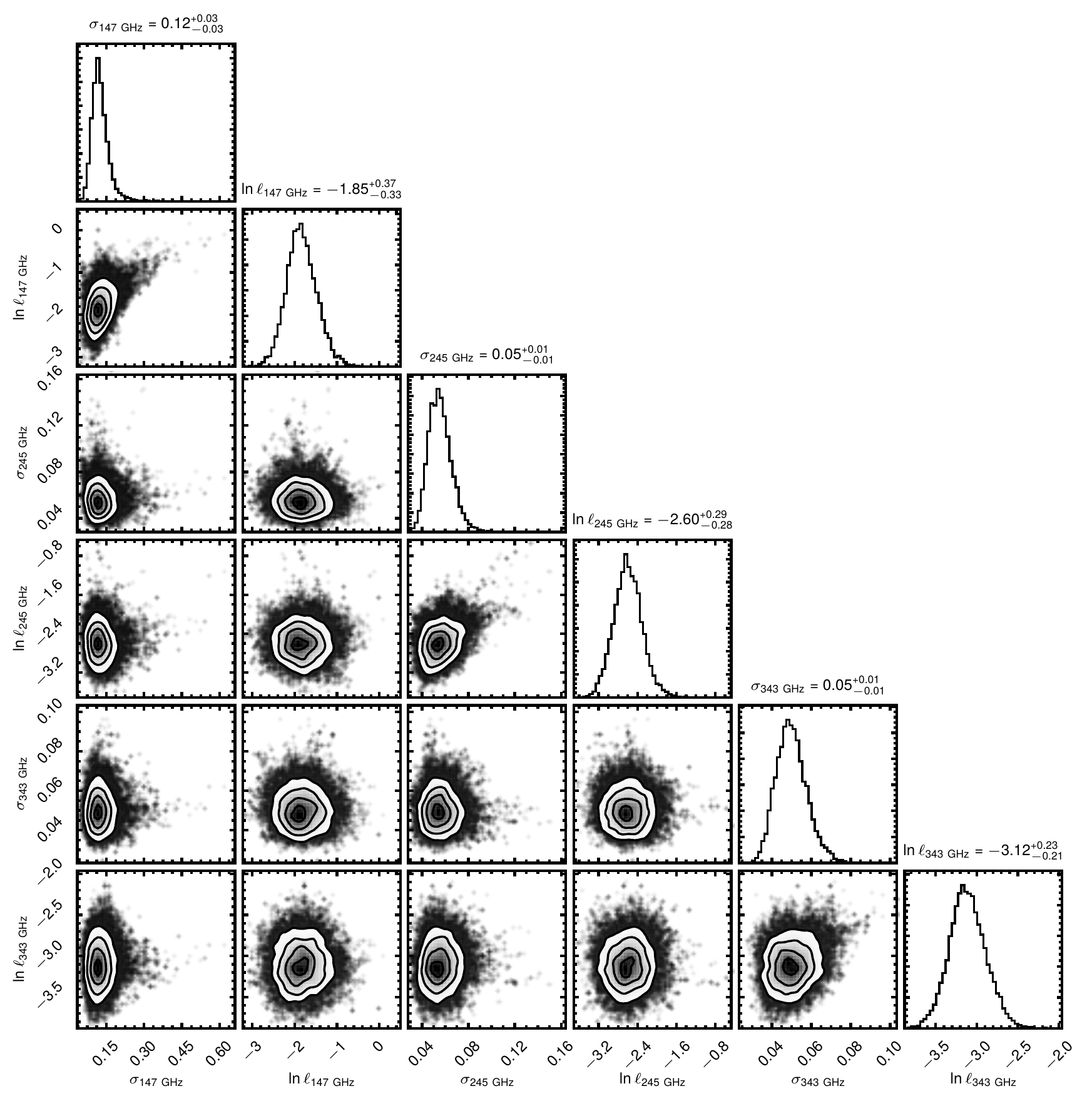}
\caption{Posterior distributions of the noise models from a representative fit at 170~au. \label{fig:app:corner_noise}}
\end{figure}

\bibliography{bibliography}

\begin{thebibliography}{}
\expandafter\ifx\csname natexlab\endcsname\relax\def\natexlab#1{#1}\fi
\providecommand{\url}[1]{\href{#1}{#1}}

\bibitem[{{ALMA Partnership} {et~al.}(2015){ALMA Partnership}, {Brogan},
  {P{\'e}rez}, {Hunter}, {Dent}, {Hales}, {Hills}, {Corder}, {Fomalont},
  {Vlahakis}, {Asaki}, {Barkats}, {Hirota}, {Hodge}, {Impellizzeri}, {Kneissl},
  {Liuzzo}, {Lucas}, {Marcelino}, {Matsushita}, {Nakanishi}, {Phillips},
  {Richards}, {Toledo}, {Aladro}, {Broguiere}, {Cortes}, {Cortes}, {Espada},
  {Galarza}, {Garcia-Appadoo}, {Guzman-Ramirez}, {Humphreys}, {Jung}, {Kameno},
  {Laing}, {Leon}, {Marconi}, {Mignano}, {Nikolic}, {Nyman}, {Radiszcz},
  {Remijan}, {Rod{\'o}n}, {Sawada}, {Takahashi}, {Tilanus}, {Vila Vilaro},
  {Watson}, {Wiklind}, {Akiyama}, {Chapillon}, {de Gregorio-Monsalvo}, {Di
  Francesco}, {Gueth}, {Kawamura}, {Lee}, {Nguyen Luong}, {Mangum}, {Pietu},
  {Sanhueza}, {Saigo}, {Takakuwa}, {Ubach}, {van Kempen}, {Wootten},
  {Castro-Carrizo}, {Francke}, {Gallardo}, {Garcia}, {Gonzalez}, {Hill},
  {Kaminski}, {Kurono}, {Liu}, {Lopez}, {Morales}, {Plarre}, {Schieven},
  {Testi}, {Videla}, {Villard}, {Andreani}, {Hibbard}, \&
  {Tatematsu}}]{ALMA_ea_2015}
{ALMA Partnership}, {Brogan}, C.~L., {P{\'e}rez}, L.~M., {et~al.} 2015, \apjl,
  808, L3

\bibitem[{{Andrews} {et~al.}(2016){Andrews}, {Wilner}, {Zhu}, {Birnstiel},
  {Carpenter}, {P{\'e}rez}, {Bai}, {{\"O}berg}, {Hughes}, {Isella}, \&
  {Ricci}}]{Andrews_ea_2016}
{Andrews}, S.~M., {Wilner}, D.~J., {Zhu}, Z., {et~al.} 2016, \apjl, 820, L40

\bibitem[{{Bai}(2015)}]{Bai_2015}
{Bai}, X.-N. 2015, \apj, 798, 84

\bibitem[{{Bai}(2017)}]{Bai_2017}
---. 2017, \apj, 845, 75

\bibitem[{{Bailer-Jones} {et~al.}(2018){Bailer-Jones}, {Rybizki}, {Fouesneau},
  {Mantelet}, \& {Andrae}}]{Bailer-Jones_ea_2018}
{Bailer-Jones}, C.~A.~L., {Rybizki}, J., {Fouesneau}, M., {Mantelet}, G., \&
  {Andrae}, R. 2018, ArXiv e-prints, arXiv:1804.10121

\bibitem[{{Balbus} \& {Hawley}(1998)}]{Balbus_Hawley_1998}
{Balbus}, S.~A., \& {Hawley}, J.~F. 1998, Reviews of Modern Physics, 70, 1

\bibitem[{{Bergin} {et~al.}(2016){Bergin}, {Du}, {Cleeves}, {Blake}, {Schwarz},
  {Visser}, \& {Zhang}}]{Bergin_ea_2016}
{Bergin}, E.~A., {Du}, F., {Cleeves}, L.~I., {et~al.} 2016, \apj, 831, 101

\bibitem[{{Bergin} {et~al.}(2013){Bergin}, {Cleeves}, {Gorti}, {Zhang},
  {Blake}, {Green}, {Andrews}, {Evans}, {Henning}, {{\"O}berg}, {Pontoppidan},
  {Qi}, {Salyk}, \& {van Dishoeck}}]{Bergin_ea_2013}
{Bergin}, E.~A., {Cleeves}, L.~I., {Gorti}, U., {et~al.} 2013, \nat, 493, 644

\bibitem[{{Boehler} {et~al.}(2017){Boehler}, {Weaver}, {Isella}, {Ricci},
  {Grady}, {Carpenter}, \& {Perez}}]{Boehler_ea_2017}
{Boehler}, Y., {Weaver}, E., {Isella}, A., {et~al.} 2017, \apj, 840, 60

\bibitem[{{Calvet} {et~al.}(2002){Calvet}, {D'Alessio}, {Hartmann}, {Wilner},
  {Walsh}, \& {Sitko}}]{Calvet_ea_2002}
{Calvet}, N., {D'Alessio}, P., {Hartmann}, L., {et~al.} 2002, \apj, 568, 1008

\bibitem[{{Cleeves} {et~al.}(2015){Cleeves}, {Bergin}, {Qi}, {Adams}, \&
  {{\"O}berg}}]{Cleeves_ea_2015a}
{Cleeves}, L.~I., {Bergin}, E.~A., {Qi}, C., {Adams}, F.~C., \& {{\"O}berg},
  K.~I. 2015, \apj, 799, 204

\bibitem[{{Cuzzi} {et~al.}(2001){Cuzzi}, {Hogan}, {Paque}, \&
  {Dobrovolskis}}]{Cuzzi_ea_2001}
{Cuzzi}, J.~N., {Hogan}, R.~C., {Paque}, J.~M., \& {Dobrovolskis}, A.~R. 2001,
  \apj, 546, 496

\bibitem[{{Czekala} {et~al.}(2017){Czekala}, {Mandel}, {Andrews}, {Dittmann},
  {Ghosh}, {Montet}, \& {Newton}}]{Czekala_ea_2017}
{Czekala}, I., {Mandel}, K.~S., {Andrews}, S.~M., {et~al.} 2017, \apj, 840, 49

\bibitem[{{Denis-Alpizar} {et~al.}(2018){Denis-Alpizar}, {Stoecklin},
  {Guilloteau}, \& {Dutrey}}]{Denis-Alpizar_ea_2018}
{Denis-Alpizar}, O., {Stoecklin}, T., {Guilloteau}, S., \& {Dutrey}, A. 2018,
  \mnras, 1112

\bibitem[{{Dipierro} {et~al.}(2018){Dipierro}, {Ricci}, {P{\'e}rez}, {Lodato},
  {Alexander}, {Laibe}, {Andrews}, {Carpenter}, {Chandler}, {Greaves}, {Hall},
  {Henning}, {Kwon}, {Linz}, {Mundy}, {Sargent}, {Tazzari}, {Testi}, \&
  {Wilner}}]{Dipierro_ea_2018}
{Dipierro}, G., {Ricci}, L., {P{\'e}rez}, L., {et~al.} 2018, \mnras, 475, 5296

\bibitem[{{Dutrey} {et~al.}(2017){Dutrey}, {Guilloteau}, {Pi{\'e}tu},
  {Chapillon}, {Wakelam}, {Di Folco}, {Stoecklin}, {Denis-Alpizar}, {Gorti},
  {Teague}, {Henning}, {Semenov}, \& {Grosso}}]{Dutrey_ea_2017}
{Dutrey}, A., {Guilloteau}, S., {Pi{\'e}tu}, V., {et~al.} 2017, ArXiv e-prints,
  arXiv:1706.02608

\bibitem[{{Fang} \& {White}(2004)}]{Fang_ea_2004}
{Fang}, T., \& {White}, M. 2004, \apj, 606, L9

\bibitem[{{Favre} {et~al.}(2013){Favre}, {Cleeves}, {Bergin}, {Qi}, \&
  {Blake}}]{Favre_ea_2013}
{Favre}, C., {Cleeves}, L.~I., {Bergin}, E.~A., {Qi}, C., \& {Blake}, G.~A.
  2013, \apjl, 776, L38

\bibitem[{{Fedele} {et~al.}(2017){Fedele}, {Carney}, {Hogerheijde}, {Walsh},
  {Miotello}, {Klaassen}, {Bruderer}, {Henning}, \& {van
  Dishoeck}}]{Fedele_ea_2017}
{Fedele}, D., {Carney}, M., {Hogerheijde}, M.~R., {et~al.} 2017, \aap, 600, A72

\bibitem[{{Flaherty} {et~al.}(2015){Flaherty}, {Hughes}, {Rosenfeld},
  {Andrews}, {Chiang}, {Simon}, {Kerzner}, \& {Wilner}}]{Flaherty_ea_2015}
{Flaherty}, K.~M., {Hughes}, A.~M., {Rosenfeld}, K.~A., {et~al.} 2015, \apj,
  813, 99

\bibitem[{{Flaherty} {et~al.}(2018){Flaherty}, {Hughes}, {Teague}, {Simon},
  {Andrews}, \& {Wilner}}]{Flaherty_ea_2018}
{Flaherty}, K.~M., {Hughes}, A.~M., {Teague}, R., {et~al.} 2018, \apj, 856, 117

\bibitem[{{Flaherty} {et~al.}(2017){Flaherty}, {Hughes}, {Rose}, {Simon}, {Qi},
  {Andrews}, {K{\'o}sp{\'a}l}, {Wilner}, {Chiang}, {Armitage}, \&
  {Bai}}]{Flaherty_ea_2017}
{Flaherty}, K.~M., {Hughes}, A.~M., {Rose}, S.~C., {et~al.} 2017, \apj, 843,
  150

\bibitem[{{Flock} {et~al.}(2017){Flock}, {Nelson}, {Turner}, {Bertrang},
  {Carrasco-Gonz{\'a}lez}, {Henning}, {Lyra}, \& {Teague}}]{Flock_ea_2017}
{Flock}, M., {Nelson}, R.~P., {Turner}, N.~J., {et~al.} 2017, \apj, 850, 131

\bibitem[{{Flock} {et~al.}(2015){Flock}, {Ruge}, {Dzyurkevich}, {Henning},
  {Klahr}, \& {Wolf}}]{Flock_ea_2015}
{Flock}, M., {Ruge}, J.~P., {Dzyurkevich}, N., {et~al.} 2015, \aap, 574, A68

\bibitem[{{Flower} \& {Watt}(1984)}]{Flower_ea_1984a}
{Flower}, D.~R., \& {Watt}, G.~D. 1984, \mnras, 209, 25

\bibitem[{{Flower} \& {Watt}(1985)}]{Flower_ea_1985b}
---. 1985, \mnras, 213, 991

\bibitem[{{Foreman-Mackey} {et~al.}(2017){Foreman-Mackey}, {Agol}, {Angus}, \&
  {Ambikasaran}}]{celerite}
{Foreman-Mackey}, D., {Agol}, E., {Angus}, R., \& {Ambikasaran}, S. 2017, AJ,
  154, 220.
\newblock \url{https://arxiv.org/abs/1703.09710}

\bibitem[{{Foreman-Mackey} {et~al.}(2013){Foreman-Mackey}, {Hogg}, {Lang}, \&
  {Goodman}}]{emcee}
{Foreman-Mackey}, D., {Hogg}, D.~W., {Lang}, D., \& {Goodman}, J. 2013, PASP,
  125, 306

\bibitem[{{Forgan} {et~al.}(2012){Forgan}, {Armitage}, \&
  {Simon}}]{Forgan_ea_2012}
{Forgan}, D., {Armitage}, P.~J., \& {Simon}, J.~B. 2012, \mnras, 426, 2419

\bibitem[{{Fromang} \& {Nelson}(2006)}]{Fromang_Nelson_2006}
{Fromang}, S., \& {Nelson}, R.~P. 2006, \aap, 457, 343

\bibitem[{{Gammie}(2001)}]{Gammie_2001}
{Gammie}, C.~F. 2001, \apj, 553, 174

\bibitem[{{Goldsmith} \& {Langer}(1999)}]{Goldsmith_Langer_1999}
{Goldsmith}, P.~F., \& {Langer}, W.~D. 1999, \apj, 517, 209

\bibitem[{{Gorti} {et~al.}(2011){Gorti}, {Hollenbach}, {Najita}, \&
  {Pascucci}}]{Gorti_ea_2011}
{Gorti}, U., {Hollenbach}, D., {Najita}, J., \& {Pascucci}, I. 2011, \apj, 735,
  90

\bibitem[{{Guilloteau} {et~al.}(2012){Guilloteau}, {Dutrey}, {Wakelam},
  {Hersant}, {Semenov}, {Chapillon}, {Henning}, \&
  {Pi{\'e}tu}}]{Guilloteau_ea_2012}
{Guilloteau}, S., {Dutrey}, A., {Wakelam}, V., {et~al.} 2012, \aap, 548, A70

\bibitem[{{Guilloteau} {et~al.}(2016){Guilloteau}, {Pi{\'e}tu}, {Chapillon},
  {Di Folco}, {Dutrey}, {Henning}, {Semenov}, {Birnstiel}, \&
  {Grosso}}]{Guilloteau_ea_2016}
{Guilloteau}, S., {Pi{\'e}tu}, V., {Chapillon}, E., {et~al.} 2016, \aap, 586,
  L1

\bibitem[{{Hendler} {et~al.}(2018){Hendler}, {Pinilla}, {Pascucci}, {Pohl},
  {Mulders}, {Henning}, {Dong}, {Clarke}, {Owen}, \&
  {Hollenbach}}]{Hendler_ea_2018}
{Hendler}, N.~P., {Pinilla}, P., {Pascucci}, I., {et~al.} 2018, \mnras, 475,
  L62

\bibitem[{{Huang} {et~al.}(2018){Huang}, {Andrews}, {Cleeves}, {{\"O}berg},
  {Wilner}, {Bai}, {Birnstiel}, {Carpenter}, {Hughes}, {Isella}, {P{\'e}rez},
  {Ricci}, \& {Zhu}}]{Huang_ea_2018}
{Huang}, J., {Andrews}, S.~M., {Cleeves}, L.~I., {et~al.} 2018, \apj, 852, 122

\bibitem[{{Hughes} {et~al.}(2011){Hughes}, {Wilner}, {Andrews}, {Qi}, \&
  {Hogerheijde}}]{Hughes_ea_2011}
{Hughes}, A.~M., {Wilner}, D.~J., {Andrews}, S.~M., {Qi}, C., \& {Hogerheijde},
  M.~R. 2011, \apj, 727, 85

\bibitem[{{Hughes} {et~al.}(2009){Hughes}, {Wilner}, {Cho}, {Marrone},
  {Lazarian}, {Andrews}, \& {Rao}}]{Hughes_ea_2009}
{Hughes}, A.~M., {Wilner}, D.~J., {Cho}, J., {et~al.} 2009, \apj, 704, 1204

\bibitem[{{Klahr} \& {Bodenheimer}(2003)}]{Klahr_Bodenheimer_2003}
{Klahr}, H.~H., \& {Bodenheimer}, P. 2003, \apj, 582, 869

\bibitem[{{Lesur} \& {Latter}(2016)}]{Lesur_Latter_2016}
{Lesur}, G. R.~J., \& {Latter}, H. 2016, \mnras, 462, 4549

\bibitem[{{Lin} \& {Youdin}(2015)}]{Lin_Youdin_2015}
{Lin}, M.-K., \& {Youdin}, A.~N. 2015, \apj, 811, 17

\bibitem[{{Loomis} {et~al.}(2018){Loomis}, {Cleeves}, {{\"O}berg}, {Aikawa},
  {Bergner}, {Furuya}, {Guzman}, \& {Walsh}}]{Loomis_ea_2018}
{Loomis}, R.~A., {Cleeves}, L.~I., {{\"O}berg}, K.~I., {et~al.} 2018, ArXiv
  e-prints, arXiv:1805.01458

\bibitem[{{Lyra} \& {Klahr}(2011)}]{Lyra_Klahr_2011}
{Lyra}, W., \& {Klahr}, H. 2011, \aap, 527, A138

\bibitem[{{Manara} {et~al.}(2016){Manara}, {Rosotti}, {Testi}, {Natta},
  {Alcal{\'a}}, {Williams}, {Ansdell}, {Miotello}, {van der Marel}, {Tazzari},
  {Carpenter}, {Guidi}, {Mathews}, {Oliveira}, {Prusti}, \& {van
  Dishoeck}}]{Manara_ea_2016}
{Manara}, C.~F., {Rosotti}, G., {Testi}, L., {et~al.} 2016, \aap, 591, L3

\bibitem[{{Marcus} {et~al.}(2015){Marcus}, {Pei}, {Jiang}, {Barranco},
  {Hassanzadeh}, \& {Lecoanet}}]{Marcus_ea_2015}
{Marcus}, P.~S., {Pei}, S., {Jiang}, C.-H., {et~al.} 2015, \apj, 808, 87

\bibitem[{{Matr{\`a}} {et~al.}(2017){Matr{\`a}}, {MacGregor}, {Kalas}, {Wyatt},
  {Kennedy}, {Wilner}, {Duchene}, {Hughes}, {Pan}, {Shannon}, {Clampin},
  {Fitzgerald}, {Graham}, {Holland}, {Pani{\'c}}, \& {Su}}]{Matra_ea_2017}
{Matr{\`a}}, L., {MacGregor}, M.~A., {Kalas}, P., {et~al.} 2017, \apj, 842, 9

\bibitem[{{Monnier} {et~al.}(2017){Monnier}, {Harries}, {Aarnio}, {Adams},
  {Andrews}, {Calvet}, {Espaillat}, {Hartmann}, {Hinkley}, {Kraus}, {McClure},
  {Oppenheimer}, {Perrin}, \& {Wilner}}]{Monnier_ea_2017}
{Monnier}, J.~D., {Harries}, T.~J., {Aarnio}, A., {et~al.} 2017, \apj, 838, 20

\bibitem[{{Mordasini} {et~al.}(2012){Mordasini}, {Alibert}, {Benz}, {Klahr}, \&
  {Henning}}]{Mordasini_ea_2012}
{Mordasini}, C., {Alibert}, Y., {Benz}, W., {Klahr}, H., \& {Henning}, T. 2012,
  \aap, 541, A97

\bibitem[{{Nelson} {et~al.}(2013){Nelson}, {Gressel}, \&
  {Umurhan}}]{Nelson_ea_2013}
{Nelson}, R.~P., {Gressel}, O., \& {Umurhan}, O.~M. 2013, \mnras, 435, 2610

\bibitem[{{P{\'e}rez} {et~al.}(2016){P{\'e}rez}, {Carpenter}, {Andrews},
  {Ricci}, {Isella}, {Linz}, {Sargent}, {Wilner}, {Henning}, {Deller},
  {Chandler}, {Dullemond}, {Lazio}, {Menten}, {Corder}, {Storm}, {Testi},
  {Tazzari}, {Kwon}, {Calvet}, {Greaves}, {Harris}, \& {Mundy}}]{Perez_ea_2016}
{P{\'e}rez}, L.~M., {Carpenter}, J.~M., {Andrews}, S.~M., {et~al.} 2016,
  Science, 353, 1519

\bibitem[{{Pohl} {et~al.}(2017){Pohl}, {Benisty}, {Pinilla}, {Ginski}, {de
  Boer}, {Avenhaus}, {Henning}, {Zurlo}, {Boccaletti}, {Augereau}, {Birnstiel},
  {Dominik}, {Facchini}, {Fedele}, {Janson}, {Keppler}, {Kral}, {Langlois},
  {Ligi}, {Maire}, {M{\'e}nard}, {Meyer}, {Pinte}, {Quanz}, {Sauvage},
  {Sezestre}, {Stolker}, {Szul{\'a}gyi}, {van Boekel}, {van der Plas},
  {Villenave}, {Baruffolo}, {Baudoz}, {Le Mignant}, {Maurel}, {Ramos}, \&
  {Weber}}]{Pohl_ea_2017}
{Pohl}, A., {Benisty}, M., {Pinilla}, P., {et~al.} 2017, \apj, 850, 52

\bibitem[{{Qi} {et~al.}(2006){Qi}, {Wilner}, {Calvet}, {Bourke}, {Blake},
  {Hogerheijde}, {Ho}, \& {Bergin}}]{Qi_ea_2006}
{Qi}, C., {Wilner}, D.~J., {Calvet}, N., {et~al.} 2006, \apj, 636, L157

\bibitem[{{Qi} {et~al.}(2004){Qi}, {Ho}, {Wilner}, {Takakuwa}, {Hirano},
  {Ohashi}, {Bourke}, {Zhang}, {Blake}, {Hogerheijde}, {Saito}, {Choi}, \&
  {Yang}}]{Qi_ea_2004}
{Qi}, C., {Ho}, P.~T.~P., {Wilner}, D.~J., {et~al.} 2004, \apjl, 616, L11

\bibitem[{{Rohlfs} \& {Wilson}(1996)}]{Rohlfs_Wilson_1996}
{Rohlfs}, K., \& {Wilson}, T.~L. 1996, {Tools of Radio Astronomy}, 127

\bibitem[{{Rosenfeld} {et~al.}(2013){Rosenfeld}, {Andrews}, {Hughes}, {Wilner},
  \& {Qi}}]{Rosenfeld_ea_2013}
{Rosenfeld}, K.~A., {Andrews}, S.~M., {Hughes}, A.~M., {Wilner}, D.~J., \&
  {Qi}, C. 2013, \apj, 774, 16

\bibitem[{{Schwarz} {et~al.}(2016){Schwarz}, {Bergin}, {Cleeves}, {Blake},
  {Zhang}, {{\"O}berg}, {van Dishoeck}, \& {Qi}}]{Schwarz_ea_2016}
{Schwarz}, K.~R., {Bergin}, E.~A., {Cleeves}, L.~I., {et~al.} 2016, \apj, 823,
  91

\bibitem[{{Shakura} \& {Sunyaev}(1973)}]{Shakura_ea_1973}
{Shakura}, N.~I., \& {Sunyaev}, R.~A. 1973, \aap, 24, 337

\bibitem[{{Simon} {et~al.}(2013){Simon}, {Bai}, {Armitage}, {Stone}, \&
  {Beckwith}}]{Simon_ea_2013}
{Simon}, J.~B., {Bai}, X.-N., {Armitage}, P.~J., {Stone}, J.~M., \& {Beckwith},
  K. 2013, \apj, 775, 73

\bibitem[{{Simon} {et~al.}(2015){Simon}, {Hughes}, {Flaherty}, {Bai}, \&
  {Armitage}}]{Simon_ea_2015}
{Simon}, J.~B., {Hughes}, A.~M., {Flaherty}, K.~M., {Bai}, X.-N., \&
  {Armitage}, P.~J. 2015, \apj, 808, 180

\bibitem[{{Teague} {et~al.}(2016){Teague}, {Guilloteau}, {Semenov}, {Henning},
  {Dutrey}, {Pi{\'e}tu}, {Birnstiel}, {Chapillon}, {Hollenbach}, \&
  {Gorti}}]{Teague_ea_2016}
{Teague}, R., {Guilloteau}, S., {Semenov}, D., {et~al.} 2016, \aap, 592, A49

\bibitem[{{Teague} {et~al.}(2017){Teague}, {Semenov}, {Gorti}, {Guilloteau},
  {Henning}, {Birnstiel}, {Dutrey}, {van Boekel}, \&
  {Chapillon}}]{Teague_ea_2017}
{Teague}, R., {Semenov}, D., {Gorti}, U., {et~al.} 2017, \apj, 835, 228

\bibitem[{{Thi} {et~al.}(2010){Thi}, {Mathews}, {M{\'e}nard}, {Woitke},
  {Meeus}, {Riviere-Marichalar}, {Pinte}, {Howard}, {Roberge}, {Sandell},
  {Pascucci}, {Riaz}, {Grady}, {Dent}, {Kamp}, {Duch{\^e}ne}, {Augereau},
  {Pantin}, {Vandenbussche}, {Tilling}, {Williams}, {Eiroa}, {Barrado},
  {Alacid}, {Andrews}, {Ardila}, {Aresu}, {Brittain}, {Ciardi}, {Danchi},
  {Fedele}, {de Gregorio-Monsalvo}, {Heras}, {Huelamo}, {Krivov}, {Lebreton},
  {Liseau}, {Martin-Zaidi}, {Mendigut{\'\i}a}, {Montesinos}, {Mora},
  {Morales-Calderon}, {Nomura}, {Phillips}, {Podio}, {Poelman}, {Ramsay},
  {Rice}, {Solano}, {Walker}, {White}, \& {Wright}}]{Thi_ea_2010}
{Thi}, W.~F., {Mathews}, G., {M{\'e}nard}, F., {et~al.} 2010, \aap, 518, L125

\bibitem[{{Trapman} {et~al.}(2017){Trapman}, {Miotello}, {Kama}, {van
  Dishoeck}, \& {Bruderer}}]{Trapman_ea_2017}
{Trapman}, L., {Miotello}, A., {Kama}, M., {van Dishoeck}, E.~F., \&
  {Bruderer}, S. 2017, \aap, 605, A69

\bibitem[{{Tsukagoshi} {et~al.}(2016){Tsukagoshi}, {Nomura}, {Muto}, {Kawabe},
  {Ishimoto}, {Kanagawa}, {Okuzumi}, {Ida}, {Walsh}, \&
  {Millar}}]{Tsukagoshi_ea_2016}
{Tsukagoshi}, T., {Nomura}, H., {Muto}, T., {et~al.} 2016, \apjl, 829, L35

\bibitem[{{van Boekel} {et~al.}(2017){van Boekel}, {Henning}, {Menu}, {de
  Boer}, {Langlois}, {M{\"u}ller}, {Avenhaus}, {Boccaletti}, {Schmid},
  {Thalmann}, {Benisty}, {Dominik}, {Ginski}, {Girard}, {Gisler}, {Lobo Gomes},
  {Menard}, {Min}, {Pavlov}, {Pohl}, {Quanz}, {Rabou}, {Roelfsema}, {Sauvage},
  {Teague}, {Wildi}, \& {Zurlo}}]{vanBoekel_ea_2017}
{van Boekel}, R., {Henning}, T., {Menu}, J., {et~al.} 2017, \apj, 837, 132

\bibitem[{{van der Tak} {et~al.}(2007){van der Tak}, {Black}, {Sch{\"o}ier},
  {Jansen}, \& {van Dishoeck}}]{vanderTak_ea_2007}
{van der Tak}, F.~F.~S., {Black}, J.~H., {Sch{\"o}ier}, F.~L., {Jansen}, D.~J.,
  \& {van Dishoeck}, E.~F. 2007, \aap, 468, 627

\bibitem[{{Walsh} {et~al.}(2017){Walsh}, {Daley}, {Facchini}, \&
  {Juh{\'a}sz}}]{Walsh_ea_2017}
{Walsh}, C., {Daley}, C., {Facchini}, S., \& {Juh{\'a}sz}, A. 2017, \aap, 607,
  A114

\bibitem[{{Weintraub} {et~al.}(1989){Weintraub}, {Masson}, \&
  {Zuckerman}}]{Weintraub_ea_1989}
{Weintraub}, D.~A., {Masson}, C.~R., \& {Zuckerman}, B. 1989, \apj, 344, 915

\bibitem[{{Yen} {et~al.}(2016){Yen}, {Koch}, {Liu}, {Puspitaningrum}, {Hirano},
  {Lee}, \& {Takakuwa}}]{Yen_ea_2016}
{Yen}, H.-W., {Koch}, P.~M., {Liu}, H.~B., {et~al.} 2016, \apj, 832, 204

\bibitem[{{Zhang} {et~al.}(2016){Zhang}, {Bergin}, {Blake}, {Cleeves},
  {Hogerheijde}, {Salinas}, \& {Schwarz}}]{Zhang_ea_2016}
{Zhang}, K., {Bergin}, E.~A., {Blake}, G.~A., {et~al.} 2016, \apjl, 818, L16

\bibitem[{{Zhang} {et~al.}(2017){Zhang}, {Bergin}, {Blake}, {Cleeves}, \&
  {Schwarz}}]{Zhang_ea_2017}
{Zhang}, K., {Bergin}, E.~A., {Blake}, G.~A., {Cleeves}, L.~I., \& {Schwarz},
  K.~R. 2017, Nature Astronomy, 1, 0130

\end{thebibliography}

\end{document}